\documentclass[11pt,english,a4paper]{article}
\pdfoutput=1
\usepackage{amsmath}
\usepackage{amssymb}
\usepackage{graphicx}
\usepackage{epsfig,jheppub}
\usepackage{subfigure}
\usepackage{float}
\usepackage{hyperref}
\usepackage{multirow}
\makeatletter

\title{Spinors Fields in Co-dimension One Braneworlds}

\author[a,b]{W. M. Mendes,}
\author[a]{G. Alencar,}
\author[a]{R. R. Landim}

\affiliation[a]{Departamento de F\'{\i}sica, Universidade Federal do Cear\'{a},
                Caixa Postal 6030, Campus do Pici, 60455-760, Fortaleza, Cear\'{a}, Brazil.\\}
\affiliation[b]{Grupo de F\'{\i}sica Teórica (GFT), Universidade Estadual do Cear\'a, Av. Dr. Silas Munguba, 1700, Campus do Itaperi, Fortaleza-CE
                CEP: 60.714.903 Cear\'{a},  Brazil.}
\emailAdd{wendelmacedo@fisica.ufc.br}
\emailAdd{geovamaciel@gmail.com}
\emailAdd{renan@fisica.ufc.br}

\abstract{
 
In this work we analyze the zero mode localization and resonances of $1/2-$spin fermions in co-dimension one Randall-Sundrum braneworld scenarios. 
We consider delta-like, domain walls and deformed domain walls membranes. Beyond the influence of the spacetime dimension $D$ we also consider
three types of couplings: (i) the standard Yukawa coupling with the scalar field and parameter $\eta_1$, (ii) a Yukawa-dilaton coupling with 
two parameters $\eta_2$ and $\lambda$ and (iii) a dilaton derivative coupling with parameter $h$. Together with the deformation parameter $s$, 
we end up with five free parameter to be considered. For the zero mode we find that the localization is dependent of $D$, because the spinorial 
representation changes when the bulk dimensionality is odd or even and must be treated separately.  For case (i) we find that in odd dimensions 
only one chirality can be localized and for even dimension a massless Dirac spinor is trapped  over the brane. In the cases (ii) and (iii) we find 
that for some values of the parameters, both chiralities can be localized in odd dimensions and for even dimensions we obtain that the massless Dirac spinor is 
trapped  over the brane. We also calculated numerically resonances for cases (ii) and (iii) by using the transfer matrix method. We find that, for 
deformed defects, the increasing of $D$ induces a shift in the peaks of resonances. For a given $\lambda$  with domain walls,  we find that the 
resonances can show up by changing the spacetime dimensionality. For example, the same case in $D=5$ do not induces resonances but when we consider 
$D=10$ one peak of resonance is found. Therefore the introduction of more dimensions, diversely from the bosonic case, can change drastically the 
zero mode and resonances in fermion fields.}

\dedicated{This paper is dedicated to my wife Gisl\^ania Mendes and my daughter Larissa Mendes (W. M. Mendes).}

\begin{document}
\maketitle

\section{Introduction}
   
     The world where we live is described by two important theories, the General Relativity and Standard Model. These theories were originally
formulated in a four-dimensional spacetime and successfully predict most of the physical quantities that we are able to measure. General Relativity is a classical theory 
that includes gravitational interactions by requiring invariance under general spacetime coordinate transformations. The Standard Model is the formulation of the quantum 
non-gravitational interactions as a consequence of a local invariance principle, generalizing the gauge invariance of electrodynamics \cite{Salvio:2007mb}. 
These theories are incompatible since a quantum theory of gravitation is not known. The unification of these theories has always been the will of physicists 
throughout history. The first model that aimed the unification of theories as electromagnetism and gravity, through a compact single extra dimension, was 
Kaluza-Klein model \cite{Kaluza:1921tu}. Although it does not exist any experimental evidence for the existence of extra dimensions, universes with higher dimension 
are important. This is the case of supersymmetric theories, for example String Theory \cite{j.polchinski1-st}. 

     In this context at the end of the last century, Lisa Randall and Raman Sundrum proposed a model that would satisfactorily solve the Higgs hierarchy problem, called  
\textit{Randall-Sundrum Model} (RSM). The setup of RSM assumes the existence of a single extra dimension and therefore spacetime is five-dimensional ($D=5$). This 
model is subdivided into two: RSM-I and RSM-II. In RSM-I the extra dimension is compactified on a circle whose upper and lower halves are identified. Formally, this 
means that we work in $S^1/\mathbb{Z}_2$ orbifold, where $S^1$ is the one-dimensional sphere (i.e., the circle) and $\mathbb{Z}_2$ is the discrete group $\{-1,1\}$. This 
construction entails two fixed points, one at the origin $\phi=0$ and one at the other extremity of the circle $\phi=\pi$. On each of these boundaries stands a 
four-dimensional world, like the one we live in. By analogy with membranes enclosing a volume, these worlds with 3+1 dimensions have been 
called 3-\textit{branes}. The RSM-I explains how a exponential hierarchy between mass scales can be generated \cite{Randall:1999ee}. The RSM-II assumes that 
the extra dimension is large (called $y$) and only one 3-brane, localized in $y=0$, exists. It shows how the gravity is confined in the 3-brane and explains how as the 
gravity in 4D emerges in the newtonian limit becoming an alternative to compactification \cite{Randall:1999vf}.

    Since in RSM-II the extra dimension is not compact all the matter fields, and not only 
gravity, must be trapped on the brane to provide a realistic model. Therefore after the conception the RSM-II, several works studied the \textit{field localization problem}.
This problem concerns the creation of a mechanisms for trapping various spin fields over the membranes. If $S^{(D)}$ is the action for the field in $D$-dimensional spacetime and the brane has dimension $(D-1)$ it is always possible to write
\begin{eqnarray}
 S^{(D)}=S^{(D-1)}\int_{-\infty}^{+\infty}dy|\phi(y)|^2=S^{(D-1)}I,
 \label{1}
\end{eqnarray}
$S^{(D-1)}$ is the effective action and $\phi(y)$ is a function of extra dimension $y$ \cite{Csaki:2000fc}. A field is said to be localized if the integral $I$ 
is finite, i.e., the function $\phi(y)$ is square integrable. Thus, it is possible to a use the Schr\"odinger picture of non-relativistic quantum mechanics to 
study brane models. 

The $D$-dimensional localizations problems with only one extra dimension are called \textit{co-dimension one} problems. The work \cite{Bajc:1999mh}
treats the field localization problem for $D=5$ in RSM-II context and shows that beyond zero mode gravity, the zero mode of scalar fields (0-spin) are localizable 
on the 3-brane. However, the same work also shows that the zero mode of gauge fields (spin 1) and the zero mode of fermionic fields (spin 1/2) are no localized over the membrane. 
In another direction localization models based in RSM-II has a singularity due to the presence of delta-like membranes, what motivated appearance of smooth brane models 
\cite{Kehagias:2000au}. In this model a scalar field  is introduced responsible for generating the brane dynamically.
In this context, beyond gravity and scalar fields, also left handed fermions (or right, but no both) are localized by a Yukawa coupling with the scalar field, but gauge fields continue non-localizable.
Recently, the problem of gauge fields localization was solved in two ways: (i) introducing a new degree of freedom in gravitation action, called dilaton\cite{Kehagias:2000au},
or (ii) introducing non-minimal coupling called \textit{geometrical coupling} \cite{Alencar:2014moa,Alencar:2015rtc,Alencar:2017dqb,Alencar:2017vqd},
where no news degrees of freedom are necessary. However, all these works considers only $D=5$ and by then nobody had studied the
conditions for localization of zero modes of these fields, except for $q$-forms \cite{Landim:2010pq,Alencar:2018cbk}. 

    The case of Spinorial fields has been considered in $D=5,6$ braneworld models \cite{Liu:2009ve,Zhao:2009ja,Li:2010dy,Almeida:2009jc,Barbosa-Cendejas:2015qaa}. In the 
    RSM-II framework Dirac's fermions are not localized, as mentioned previously. Thus, according \cite{Jackiw:1975fn} the zero mode of left handed fermions is localized 
    when a interaction of fermions with a scalar field $\phi\overline{\Psi}\Psi$  is introduced. In smooth RSM-II, zero mode of chiral massless spinorial field is 
    localized \cite{Kehagias:2000au}. When the dilation is add the spinoral fields are localized, choosing rightly dilaton coupling constant. The authors of the paper 
    \cite{Liu:2013kxz} showed that a Yukawa potential can localize fermions because the coupling between the fermion and the background scalar field is an odd function of 
    extra dimension. However, the same authors showed also that, if the scalar is an even function of the extra dimension, this mechanism does not work anymore, and we need 
    to introduce a new localization mechanism. It is worth mentioning that in \cite{Li:2017dkw} the authors showed that geometrical derivative coupling gives 
    localized fermions on the branes. We should point that the above mentioned works treat the localization problem only in in $D=5,6$.

   In another direction, even if the zero mode is not localized it is important to analyze the possible appearance of unstable massive modes. For this we 
    need to find the Schr\"odinger like equation that drives the massive modes. Therefore, as mentioned previously, our problem can be addressed as if we were dealing with an one dimensional quantum mechanical problem with potential given by $U(z)$, as in the Refs. \cite{Landim:2011ki, Landim:2011ts, Alencar:2012en}. The unstable modes can be found by looking for resonances of the effective potential. The potentials in this work satisfy the condition
\begin{eqnarray}
 \lim _{z \rightarrow \pm\infty}U(z)=0,
\end{eqnarray}
and therefore, according to elementary quantum mechanics, does not exist a discrete spectrum for $m>0$. Also, the wave function is not normalizable because any solution
for positive $m$ is a plane wave asymptotically. The conclusion is that the zero mode is the only possible eigenstate of the Schr\"odinger like equation 
localizable on the membrane.

    The goal of this paper is to study the localization of zero modes and resonances of fermions in co-dimension one braneworlds generated by delta-like branes as well 
    as smooth versions of it.  For this we consider Yukawa, dilaton and dilaton derivative couplings. This work is organizaded as follows: in Sect. \ref{Bulk} we review 
    the main aspects of $(D-2)$-Branes in $D-$dimensional Randall-Sundrum models. In Sect. 
\ref{Spinors} we review the constrution spinorial representation in arbitrary dimensions, in Sect. \ref{Localization} we discussed the zero mode fermions localization
and in Sect. \ref{Resonances} we show and discussed the arbitrary case of resonance pattern in massive modes. Finally, in Sect. \ref{End}, some final remarks.

\section{Review of Randall-Sundrum Model}
\label{Bulk}

In this section we will review some aspects of co-dimension one Randall-Sundrum braneworld scenarios. We will consider delta-like, domain walls and deformed domain 
walls membranes.

\subsection{The Delta-Like Brane}

    The thin brane model can be defined by action \cite{Kim:2003pc}
\begin{eqnarray}
 S=2M^p\int d^Dx\sqrt{-g}(R-\Lambda)-V\int\sqrt{-g^{(B)}}d^{D-1}x,
 \end{eqnarray}
where $p\equiv D-2$, $M^p$ is fundamental scale of energy of theory, $\Lambda$ is a $D$-dimensional cosmological constant, $V$ is tension of the brane and $g^{(B)}$ 
is the induced metric determinant on the $p$-brane, which obeys the boundary condition $g_{\mu\nu}(x,y=0)=g_{\mu\nu}^{(B)}(x).$ 

   The ansatz for the space-time metric in this setup, is
\begin{eqnarray}
 ds^2=e^{2A(y)}\eta_{\mu\nu}dx^\mu dx^\nu+dy^2,
 \label{dsRS}
\end{eqnarray}
where $\eta_{\mu\nu}=\mbox{diag}(-1,1,...,1)$ is the metric of $p$-brane. The Einstein equations for this model are
\begin{eqnarray}
 pA''(y)+\frac{p(p+1)}{2}A'^2(y)+\Lambda&=&\frac{V}{4M^p}\delta(y),\\ \label{EE1}
         \frac{p(p+1)}{2}A'^2(y)+\Lambda&=&0. \label{EE2}
\end{eqnarray} 
The solution for the function $A(y)$ is $A(y)=-k_p|y|$, where $k_p\equiv \sqrt{-2\Lambda/p(p+1)}$. As we can see the solution respects the orbifold symmetry, i.e.,
invariance under the transformation $y\rightarrow -y$. Therefore the interval (\ref{dsRS}) becomes \cite{Kim:2003pc}
\begin{eqnarray}
 ds^2=e^{-2k_p|y|}\eta_{\mu\nu}dx^\mu dx^\nu+dy^2.
\end{eqnarray}
In this way the interval is determined, specifically, the bulk cosmological constant $\Lambda$. Note that the RSM-II consist of one flat $p$-branes imbedded
discontinuously in a larger space. In the next subsection we will consider a version smooth of the RSM-II.

\subsection{The Thick Brane Case}
\label{Thick brane}

   In the thick brane case we consider branes as topological defects, in particular domain wall. This is represented by a real scalar field that depends only on the 
extra dimension. The ansatz for the metric is the same as (\ref{dsRS}). But the action for this model is
\begin{eqnarray}
 S=\int d^{D-1}x\int dy\sqrt{-g}\left[2M^pR-\frac{1}{2}\partial_M\phi\partial^M\phi-V(\phi)\right],
 \label{Sphi}
\end{eqnarray}
where $\phi$ is a real scalar field that generates the $p$-brane and $V(\phi)$ introduces bounce like solutions. The energy-momentum tensor of real scalar field is
\cite{Nakahara:2003nw}
\begin{eqnarray}
  T_{MN}=\partial_M\phi\partial_N\phi-g_{MN}\left[\frac{1}{2}\partial_A\phi\partial^A\phi+V(\phi)\right],
 \end{eqnarray}
and equations of motion are given by
\begin{eqnarray}
 A''(y)+\frac{(p+1)}{2}A'^2(y)&=&-\frac{1}{4pM^p}\left[\frac{1}{2}\phi'^2+V(\phi)\right], \label{E1} \\
        \frac{(p+1)}{2}A'^2(y)&=&\frac{1}{4pM^p}\left[\frac{1}{2}\phi'^2-V(\phi)\right], \label{E2}  \\
      \phi''+[(p+1)A'(y)]\phi'&=&\frac{\partial V}{\partial\phi}.
\end{eqnarray}
The solution of these equations can be obtained through the superpotential method, according to \cite{Kehagias:2000au, Csaki:2000fc, Skenderis:1999mm}, introducing the superpotencial
$W(\phi)$ by $\displaystyle \phi'=\frac{\partial W}{\partial\phi}$. Therefore, choosing $\phi(y)=a\tanh(by)$ the superpotencial $W(\phi)$ is given by 
\begin{eqnarray}
  W(\phi)=ab\phi\left(1-\frac{\phi^2}{3a^2}\right).
\end{eqnarray}
After determining the superpotential $W(\phi)$, the scalar potential $V(\phi)$ can be written as
\begin{eqnarray}
 V(\phi)&=&\frac{1}{2}\left(\frac{\partial W}{\partial\phi}\right)^2-\frac{(p+1)}{8pM^p}W^2.
\end{eqnarray}
Thus, the superpotential definition leads to
\begin{eqnarray}
 \quad A'(y)=-\frac{1}{4pM^p}W.
\end{eqnarray}
Finally we get
\begin{eqnarray}
A(y)=-\beta_p\left[\ln\cosh^2(by)+\frac{1}{2}\tanh^2(by)\right], 
 \label{A}
\end{eqnarray}
where $\beta_p\equiv a^2/4pM^p$. Note that just as in the of RSM-II case, here $A(y)$ also represents a localized and smooth metric warp factor. We will consider $a=1$ 
and $M=1$ for resonances analyzes of 1/2-spin fields. In the next subsection we will consider the braneworld scenario with a new degree of freedom: the dilaton.

\subsection{Dilaton Coupling}
\label{Dilaton}
   
    As already mentioned in the introduction, the dilaton field is commonly used in localization of gauge fields in $D=5$, for example see \cite{Kehagias:2000au,
Cruz:2010zz, Landim:2010pq, Alencar:2010hs}. This new degree of freedom is represented for $\pi$. We propose the following action for this setup:  
\begin{eqnarray}
 S=\int d^Dx\sqrt{-g}\left[2M^pR-\frac{1}{2}(\partial_M\phi\partial^M\phi+\partial_M\pi\partial^M\pi)+V(\phi,\pi)\right].
 \label{Sphipi}
\end{eqnarray}
The ansatz for the space-time interval with dilaton coupling is
\begin{eqnarray}
 ds^2=e^{2A(y)}\eta_{\mu\nu}dx^\mu dx^\nu+e^{2B(y)}dy^2.
\end{eqnarray}
The jacobian for this case is $\sqrt{-g}=e^{(p+1)A(y)+B(y)}$, that, as expected, depends on the dimensionality of space-time. The equations of motion resulting from
the action (\ref{Sphipi}) are
\begin{eqnarray}
 A''(y)+\frac{(p+1)}{2}A'^2(y)-A'(y)B'(y)&=&-\frac{1}{4pM^p}\left[\frac{1}{2}\phi'^2+\frac{1}{2}\pi'^2+e^{2B(y)}V(\phi,\pi)\right], \\
 \frac{(p+1)}{2}A'^2(y)&=&\frac{1}{4pM^p}\left[\frac{1}{2}\phi'^2+\frac{1}{2}\pi'^2-e^{2B(y)}V(\phi,\pi)\right],
 \label{DEOM1}
\end{eqnarray}
\begin{eqnarray}
 \phi''(y)+[(p+1)A'(y)-B'(y)]\phi'(y)&=&\frac{\partial V}{\partial\phi},\\
 \pi''(y)+[(p+1)A'(y)-B'(y)]\pi'(y)&=&\frac{\partial V}{\partial\pi}.
 \label{DEOM2}
\end{eqnarray}
We use again the superpotencial method with
\begin{eqnarray}
 V(\phi,\pi)=\exp{\left(\sqrt{\frac{r}{pM^p}}\pi\right)}\left[\frac{1}{2}\left(\frac{\partial W}{\partial\phi}\right)^2-\frac{(p+1-r)}{8pM^p}W^2\right],
 \label{Vphipi}
\end{eqnarray}
where $r$ is a real positive constant. The solutions for the above system of equations are given by (\ref{A}) and 

\begin{eqnarray} 
 B(y)=rA(y)=-\frac{1}{2}\sqrt{\frac{r}{pM^p}}\pi(y).
 \label{Br}
\end{eqnarray} 

    Now, according \cite{Fu:2011pu} we will investigate the values that the constant $r$ can assume. For this, we will analyze the energy density of the system $T_{00}(y)$.
 With the solution \label{B}, it is given by
\begin{eqnarray}
 T_{00}(y)=-4pM^p\left[A''(y)+\left(\frac{p+1-2r}{2}\right)A'^2(y)\right]e^{2(1-r)A(y)}.
\end{eqnarray}
Asymptotically, when $y\rightarrow \pm\infty$, we have $A(y)\rightarrow -2\beta_p|y|$, and therefore the tensor $T_{00}$ becomes
\begin{eqnarray}
 T_{00}(y)\rightarrow 8bpM^p\beta_p\left[2\delta(y)-(p+1-2r)b\right]e^{-2(1-r)b\beta_p|y|},
\end{eqnarray}
where $\delta(y)$ is the Dirac's delta. Therefore the behavior of $T_{00}$ when $y\rightarrow \pm\infty$ is
\begin{eqnarray}
 T_{00}(|y|\rightarrow\infty) &=& \left\{
                                        \begin{array}{ll}
                                            0, ~~~~~~~~~~~~~~  & 0<r<1    \\
                                         -8b^2pM^p\beta_p(p-1), ~~~~~~ &r=1 \\
                                          -\infty, ~~~~~~~~~~~~ & 1<r<(p+1)/2 \\
                                            0, ~~~~~~~~~~~~~~~  & r=(p+1)/2>1   \\
                                           \infty,~~~~~~~~~~~~~ & r>(p+1)/2
                                         \end{array}. \right.
\label{T00}
\end{eqnarray}
In the Sect. \ref{Localization} we will mainly discuss the effect of the parameter $r$ on the localization of spinorial field.

\subsection{Deformed Brane}

    The last case of our interest, is the deformed brane, approached in \cite{Landim:2011ts}. The deformation method is based in modifications of the potential of 
models containing solitons in order to produce new and unexpected solutions \cite{Bazeia:2002xg}. Now we introduce the deformation parameter, called $s$, that 
controls the kind of deforming topological defect, in order to simulate different classes of branes. For this case the interval can be written as
\begin{eqnarray}
 ds^2=e^{2A_s(y)}\eta_{\mu\nu}dx^\mu dx^\nu+e^{2B_s(y)}dy^2.
\end{eqnarray}
As in subsections \ref{Thick brane} and \ref{Dilaton} we will use the superpotential method to solve the system of equations
\begin{eqnarray}
 V_s(\phi_s,\pi_s)=\exp{\left(\sqrt{\frac{r}{pM^p}}\pi_s\right)}\left[\frac{1}{2}\left(\frac{\partial W_s}{\partial\phi_s}\right)^2-\frac{(p+1-r)}{8pM^p}W_s^2\right].
\end{eqnarray}
The superpotential chosen is
\begin{eqnarray}
 W_s(\phi_s)=b\phi_s^2\left[\frac{s}{2s-1}\left(\frac{a}{\phi_s}\right)^{1/s}-\frac{s}{2s+1}\left(\frac{\phi_s}{a}\right)^{1/s}\right],
\end{eqnarray}
and 
\begin{eqnarray}
\phi_s(y)&=&a\tanh^s\left(\frac{by}{s}\right),\\
 A_s(y)&=&-\beta_{s,p}\tanh^{2s}\left(\frac{by}{s}\right)-\frac{2s\beta_{p,s}}{2s-1}\left[\ln\cosh\left(\frac{by}{s}\right)-
                                                                 \sum_{n=1}^{s-1}\frac{1}{2n}\tanh^{2n}\left(\frac{by}{s}\right)\right], 
 \label{As}
\end{eqnarray}
where $\beta_{p,s}\equiv \frac{a^2}{4pM^p}\frac{s}{2s+1}$. Note that, for $s=1$, the undeformed solutions are reproduced. In the next sections, we will use the above 
backgrounds to study the localization and resonances of spinor fields in arbitrary dimensions.

\section{Dimensional Reduction and Mass Equation for Fermions in Arbitrary Dimensions}
\label{Spinors}

    As mentioned in the introduction, the mass spectrum of the fields in RSM-II is driven by a Schr\"odinger like equation. In this section we derive this equation for 
an arbitrary space-time dimension.  We will see that the cases with odd and even dimensions needs to treated separately. As a starting point, we will consider the 
$D-$dimensional action for massless Dirac fermion in a curved background given by
\begin{eqnarray}
 S_{1/2}=\int d^Dx\sqrt{-g}\mbox{ }\overline{\Psi}[\Gamma^MD_M-F(\phi,\pi)-\Gamma^M\partial_MH(\phi,\pi)\Gamma]\Psi,
 \label{S1/2}
\end{eqnarray}
where $F(\phi,\pi)$, $H(\phi,\pi)$ are interaction terms which involves the scalar field $\phi$ and dilation $\pi$ and $D_M=\partial_M+\omega_M$ is the covariant derivate
and $\omega_M$ is the spin connection. The equation of motion resulting from the action (\ref{S1/2}) is
\begin{eqnarray}
\left[\Gamma^MD_M-F(\phi,\pi)-\Gamma^M\partial_MH(\phi,\pi)\Gamma\right]\Psi(x,y)=0.
\label{DiracEOM}
\end{eqnarray}
In a curved space-time, the Clifford algebra reads $\{ \Gamma^M,\Gamma^N\}=2g^{AB}$, where $g^{MN}=e_A{}^Me_B{}^N\eta^{AB}$ is the curved space metric and 
$e_A{}^M$ are the vierbeins. The index $A,B,C,...,J$ are Lorentz index and $L,M,N,...,Z$ are curved space  index. The spin connection is defined by
\begin{eqnarray}
 \omega_M=\frac{1}{4}e^A{}_L(\partial_Me^{BL}+\Gamma^L{}_{MN}e^{BN})\Gamma_A\Gamma_B.
 \label{Cspin}
\end{eqnarray}
The nonzero spin connection components for the metric (\ref{dsRS}) are
\begin{eqnarray}
 \omega_\mu=\frac{1}{2}A'(y)e^{A(y)-B(y)}\Gamma_\mu\Gamma_y,
\end{eqnarray}
where greek index represents the coordinates in the brane. Therefore, the equation (\ref{DiracEOM}) can be written as
\begin{eqnarray}
\left(\gamma^\mu\partial_\mu+e^{A(y)-B(y)}\left[\partial_y+\frac{(p+1)}{2}A'(y)\right]\Gamma_y-e^{A(y)}\left[F(\phi,\pi)+\partial_yH(\phi,\pi)\right]\right)\Psi(x,y)=0.
 \label{eqDiracG}
\end{eqnarray}
Now, we need a better understanding of fermion representations in arbitrary dimensions since the behavior of this fields are different in odd or even dimensions. 
This is the topic of next section.

\subsection{Review of Spinors in Arbitrary Dimensions}
\label{Representation}
    
    The spinorial representations of $SO(1,D-1)$ group changes depending on the dimensionality of space-time \cite{Cartan}. There is a very instructive way to build 
the 1/2 spin representations in arbitrary dimensions \cite{j.polchinski2-st}, which is very common in String Theory. The Clifford algebra, the starting point to 
build spinorial representation of Lorentz group, is given by
\begin{eqnarray}
 \{ \Gamma^A,\Gamma^B\}=2\eta^{AB},
 \label{Clifford D}
\end{eqnarray}
where $\Gamma^A$ are Dirac matrices  and $\eta_{AB}=\mbox{diag}(-1,1,1...,1)$ is the  Minkowski metric. We first consider  that the dimensionality $D$ of space-time is 
even, i.e., $D=2k+2$, where $k$ is a positive integer number. It is known that there is a relationship between the rotation and the Lorentz groups. The spin idea is 
introduced through the generators 
\begin{eqnarray}
 \Sigma^{AB}\equiv -\frac{i}{4}[\Gamma^A,\Gamma^B],
\end{eqnarray}
satisfying the Lie algebra of Lorentz group. The Dirac representation is not irreducible, and can be broken in two disjunct subspaces, called Weyl 
representation. In this representation we define the matrix 
\begin{eqnarray}
 \Gamma\equiv i^{-k}\prod_{l=0}^{D-1}\Gamma^l=i^{-k}\Gamma^0\Gamma^1...\Gamma^{D-1},
 \label{Gamma}
\end{eqnarray}
which has the properties
\begin{eqnarray}
 (\Gamma)^2=\mathbf{1}, \quad  \{\Gamma,\Gamma^A\}=0, \quad [\Gamma,\Sigma^{AB}]=0.
 \label{properties}
\end{eqnarray}
The eigenvalues of $\Gamma$, called chirality, are $\pm1$. The $2^k$ states with $\Gamma$ eigenvalue $+1$ form a Weyl representation of the Lorentz algebra, 
and the $2^k$ states with eigenvalue $-1$ form a second, inequivalent, Weyl representation.

 In two-dimensions, one can take the Pauli matrices $\sigma^1$, $\sigma^2$ and $\sigma^3$ as gamma matrices, i.e., $\Gamma^0=i\sigma^2$ and $\Gamma^1=\sigma^1$.
Also $\Gamma=\sigma^3$. In $D=2k+2$ dimensions, the gamma matrices can be written as \cite{Park-2005}
\begin{eqnarray}
 \Gamma^{(\alpha)}=\gamma^{(\alpha)}\otimes\sigma^1, \quad \Gamma^{D-2}=\gamma\otimes\sigma^1, \quad \Gamma^{D-1}=\mathbf{1}\otimes\sigma^3,  
\end{eqnarray}
where $(\alpha)=0,1,2,..., D-3$ and $\gamma^{(\alpha)}$, $\gamma$ and $\mathbf{1}$ are $2k\times2k$ Dirac matrices, chirality and identity in $D-2$ dimensions.

For an odd dimension, i.e, $D=2k+3$, the matrix $\Gamma=\Gamma^D$ is included in the Clifford algebra, due to the (\ref{properties}), and does not exists chirality
in this dimensionality. Now, we have the necessary mathematical tools for perform the dimensional reduction of the mass Dirac equation (\ref{eqDiracG}).

\subsection{Reduction of Mass Equation in Odd Dimensions}

    By using the results of the last section now we can consider the dimensional reduction for the cases with odd dimensions. In this case we perform the 
decomposition
\begin{eqnarray}
\Psi(x,y)=e^{-\frac{(p+1)}{2}A(y)}\sum_{n}\left[\psi_{n+}(x)f_{n+}(y)+\psi_{n-}(x)f_{n-}(y)\right],
\label{decomposition}
\end{eqnarray}
where we defined $\gamma\psi_{n+}(x)=\psi_{n+}(x)$ and $\gamma\psi_{n-}(x)=-\psi_{n-}(x)$ as the left and right-handed fermion fields in $D-1$ dimensions, 
respectively. 
The equations for massive field $D-1$ dimensions are 
\begin{eqnarray}
 \gamma^\alpha\partial_\alpha\psi_{n+}(x)&=&m_n\psi_{n-}(x), \label{L}\\
 \gamma^\alpha\partial_\alpha\psi_{n-}(x)&=&m_n\psi_{n+}(x). \label{R}
\end{eqnarray}
Thus, replacing (\ref{decomposition}) in  (\ref{eqDiracG}), the equation of motion for extra dimension becomes
\begin{eqnarray}
 \left\{\partial_y+e^{B(y)}\left[F(\phi,\pi)+\partial_yH(\phi,\pi)\right]\right\}f_{n-}(y)&=&m_ne^{B(y)-A(y)}f_{n+}(y), \label{Eqodd1} \\
 \left\{\partial_y-e^{B(y)}\left[F(\phi,\pi)+\partial_yH(\phi,\pi)\right]\right\}f_{n+}(y)&=&-m_ne^{B(y)-A(y)}f_{n-}(y). \label{Eqodd2}
\end{eqnarray}
These equations can be transformed in equations of the type
\begin{eqnarray}
 \left[-\frac{d^2}{dy^2}+P'(y)\frac{d}{dy}+V(y)\right]\Phi(y)=m^2Q(y)\Phi(y).
\end{eqnarray}
The best way to study this equation is to transform it in a Schr\"odinger like equation
\begin{eqnarray}
 \left[-\frac{d^2}{dz^2}+U(z)\right]\bar{\Phi}(z)=m^2\bar{\Phi}(z), 
 \label{Schoedinger equation}
\end{eqnarray} 
\begin{eqnarray}
 \frac{dz}{dy}=f(y), \quad \Phi(y)=\Omega(y)\bar{\Phi}(z),
\end{eqnarray}
 through the transformations \cite{Landim:2011ki} 
\begin{eqnarray}
f(y)=\sqrt{Q(y)}, \quad \Omega(y)=\exp\left[\frac{P(y)}{2}\right]Q(y)^{-1/4},
\label{Transformations}
\end{eqnarray}
and
\begin{eqnarray}
 U(z)=\frac{V(y)}{f(y)^2}+\frac{P'(y)\Omega'(y)-\Omega''(y)}{\Omega(y)f(y)^2},
 \label{U(z)}
\end{eqnarray}
where the prime is a derivative with respect to $y$. The effective potentials are
\begin{eqnarray}
 U_\pm(z)=e^{2A(y)}\left[F(\phi,\pi)+\partial_yH(\phi,\pi)\right]^2\pm\frac{e^{-B(y)}}{2}\partial_y\{e^{2A(y)}\left[F(\phi,\pi)+\partial_yH(\phi,\pi)\right]\},
 \label{potentiais+-}
\end{eqnarray}
where the the changing of variables is defined by $dz/dy=e^{B(y)-A(y)}$.

\subsection{Reduction of Mass Equation in Even Dimensions}

For the case where the dimensionality is even, the separation of variables is 
\begin{eqnarray}
 \Psi(x,y)=e^{-\frac{(p+1)}{2}A(y)}\sum_n\psi_n(x)\otimes\xi_n(y),
 \label{SolPar}
\end{eqnarray}
where $\psi(x)$ is a $2^k$-dimensional Dirac's spinor and $\xi_n(y)$ is a two dimensional spinor. Replacing $\Gamma_y=\mathbf{1}\otimes\sigma^3$ the Dirac 
operator $\Gamma^MD_M\Psi(x,y)$ becomes
\begin{eqnarray}
 \Gamma^MD_M\Psi(x,y)=e^{-\frac{(p-3)}{2}A(y)}\sum_n\psi_n(x)\otimes\left[m_n\sigma^1+e^{A(y)-B(y)}\sigma^3\partial_y\right]\xi_n(y).
 \label{DiracOperator}
\end{eqnarray}

So replacing (\ref{SolPar}) in  (\ref{eqDiracG}), the equation for massive modes in the extra dimension is
\begin{eqnarray}
  \frac{d\xi_n(y)}{dy}-e^{B(y)}\left\{\left[F(\phi,\pi)+\partial_yH(\phi,\pi)\right]\sigma^3-m_ne^{-A(y)}i\sigma^2\right\}\xi_n(y)=0.
  \label{Eqxi}
\end{eqnarray}
The matrix $\sigma^3$ is the chirality matrix in two dimensions, where $\Gamma^0=i\sigma^2$ and $\Gamma^1=\sigma^1$ are Dirac matrices. So we can write the two 
dimensional spinor $\xi_n(y)$ as
\begin{eqnarray}
 \xi_n(y)=\left[\begin{array}{c}
           \xi_{n+}(y)\\
           \xi_{n-}(y)\\
                \end{array}\right],
                \label{xin}
\end{eqnarray}
where  $\xi_{n+}(y)$ and $\xi_{n-}(y)$ are complex functions. Therefore replacing (\ref{xin}) in the equation (\ref{Eqxi}), we obtain the following differential equations 
\begin{eqnarray}
 \left\{\partial_y+e^{B(y)}\left[F(\phi,\pi)+\partial_yH(\phi,\pi)\right]\right\}\xi_{n-}(y)&=&m_ne^{B(y)-A(y)}\xi_{n+}(y), \label{Eqeven1}\\
 \left\{\partial_y-e^{B(y)}\left[F(\phi,\pi)+\partial_yH(\phi,\pi)\right]\right\}\xi_{n+}(y)&=&-m_ne^{B(y)-A(y)}\xi_{n-}(y). \label{Eqeven2}
\end{eqnarray}
These equations are identical to those related the odd dimensionality, i.e., equations (\ref{Eqodd1}) and (\ref{Eqodd2}). Therefore, the potentials and the 
Schr\"odinger's equation are the same however, here, it is possible choose $\xi_{n+}(y)\neq 0$ and $\xi_{n-}(y)=0$ or $\xi_{n+}(y)=0$ and $\xi_{n-}(y)\neq 0$. This 
choice implies that not exists massive modes in the membrane. In even dimensions it is possible because this procedure is equivalent to take Weyl spinors 
\begin{eqnarray}
 \Psi_+(x,y)=\sum_n\psi_n(x)\otimes\left[\begin{array}{c}
                                    \xi_{n+}(y)\\
                                         0     \\
                                         \end{array}\right], \quad \Psi_-(x,y)=\sum_n\psi_n(x)\otimes\left[\begin{array}{c}
                                                                                                            0     \\
                                                                                                       \xi_{n-}(y)\\
                                                                                                            \end{array}\right],
\end{eqnarray}
from the beginning in the equations of motion. This procedure is not true for odd dimensions because in this case, according to Sect. \ref{Representation}, chirality is not defined. 
These results will be used to study the zero mode localization and resonances of fermions in arbitrary space-time dimensions.

\section{The Zero Mode Localization}
\label{Localization}

   In this section we will analyze the localization of zero modes over the membrane. The condition for localization is that we can obtain a well defined $(D-1)-$dimensional action. 
Therefore we look for solutions such that the integral (\ref{1}) is finite, which is like the square integrable condition of quantum mechanics. Together with transformations 
(\ref{Transformations}) we obtain a standard Schr\"odinger-like equation problem. According to the subsection \ref{Representation}, the Dirac matrix representation change when the 
bulk dimension is odd or even. First we will attack the zero mode localization problem when the dimensionality is odd, after the case where it is even. 

\subsection{General Case}
\label{General case}

 For the case that the dimensionality is odd, the equations (\ref{Eqodd1}) and (\ref{Eqodd2}) for the zero mode reduces to
\begin{eqnarray}
 f'_{\pm0}(y)\mp e^{B(y)}\left[F(\phi,\pi)+\partial_yH(\phi,\pi)\right]f_{\pm0}(y)=0.
\end{eqnarray}
The solutions to these equations are
\begin{eqnarray}
 f_{\pm0}(y)=C_\pm\exp\left\{\pm\int_y dy'e^{B(y')}\left[F(\phi(y'),\pi(y'))+\partial_yH(\phi(y'),\pi(y'))\right]\right\},
 \label{f(y)}
\end{eqnarray}
where the positive signal refers to right, the negative signal to left handed spinors and $C_{\pm}$ are arbitrary constants. We note that, in this scenario, the interaction 
term is fundamental for the localization of spinorial fields. Replacing (\ref{decomposition}) in the action (\ref{S1/2}) for the zero mode, the effective $(D-1)$-dimensional 
action becomes
\begin{eqnarray}
 S_{1/2}=\int d^{D-1}x\mbox{ }\overline{\psi}_{0+}\gamma^\mu\partial_\mu\psi_{0+}\left[\int^{+\infty}_{-\infty}dye^{-A(y)+B(y)}|f_{0+}(y)|^2\right]+\nonumber\\
         +\int d^{D-1}x\mbox{ }\overline{\psi}_{0-}\gamma^\mu\partial_\mu\psi_{0-}\left[\int^{+\infty}_{-\infty}dye^{-A(y)+B(y)}|f_{0-}(y)|^2\right]
\end{eqnarray}
The localization of the zero mode will depend on the finitude of the integrals 
\begin{eqnarray}
 I_{odd\pm}&=&\int^{+\infty}_{-\infty}dye^{-A(y)+B(y)}|f_{0\pm}(y)|^2
 \label{intloc1}
\end{eqnarray}
which, by using the solution (\ref{f(y)}), becomes
\begin{eqnarray}
 I_{odd\pm}=\int^{+\infty}_{-\infty}dy\exp{\left\{-A(y)+B(y)\pm2\int_y dy'e^{B(y')}\left[F(\phi,\pi)+\partial_yH(\phi,\pi)\right]\right\}}.
\label{Ioddpm}
\end{eqnarray}

    Now we turn our attention to even dimensions. In this case the equations (\ref{Eqeven1}) and (\ref{Eqeven2}) for $m=0$ becomes
\begin{eqnarray}
 \xi'_{0\pm}(y)\mp e^{B(y)}\left[F(\phi,\pi)+\partial_yH(\phi,\pi)\right]\xi_{0\pm}(y)=0,
\end{eqnarray}
with solution
\begin{eqnarray}
 \xi_{0\pm}(y)=\bar{\xi}_\pm\exp\left\{\pm\int_y dy'e^{B(y')}\left[F(\phi(y'),\pi(y'))+\partial_yH(\phi(y'),\pi(y'))\right]\right\},
 \label{xi(y)}
\end{eqnarray}
where $\bar{\xi}_\pm$ are arbitrary complex constants. Replacing (\ref{SolPar}) in the action (\ref{S1/2}), the effective $(D-1)-$dimensional action for the zero mode is
\begin{eqnarray}
 S_{1/2}=\int d^{D-1}x\mbox{ }\overline{\psi}_0\gamma^\mu\partial_\mu\psi_0\int^{+\infty}_{-\infty}dye^{-A(y)+B(y)}\left[|\xi_{0+}(y)|^2+|\xi_{0-}(y)|^2\right].
 \label{S-even}
\end{eqnarray}
Similarly to the previous case, we must analyze the convergence of integral 
\begin{eqnarray}
 I_{even}=\int^{+\infty}_{-\infty}dye^{-A(y)+B(y)}\left[|\xi_{0+}(y)|^2+|\xi_{0-}(y)|^2\right].
\end{eqnarray}
Therefore, as previously case, replacing the equation (\ref{xi(y)}) in the integral $I_{even}$, it can be written as
\begin{eqnarray}
 I_{even}=I_{odd+}+I_{odd-} \quad .
\label{Ieven}
\end{eqnarray}
According to equation (\ref{Ieven}), the Dirac spinor $\psi_0(x)$ is localized only if both $I_{odd+}$ and $I_{odd-}$ are finite. In many cases, this condition
is not obeyed simultaneously, compromising the localization mechanism. This apparent problem can be solved, only for zero mode, choosing appropriately the functions
$\xi_{0+}(y)$ and $\xi_{0+}(y)$ such that they continue to satisfy equations (\ref{Eqeven1}) and (\ref{Eqeven2}). We will choose $\xi_{0+}(y)\neq 0$ and $\xi_{0-}(y)=0$ 
if the integral $I_{odd-}$ is divergent, and therefore the equation (\ref{Ieven}) becomes $I_{even}=I_{odd+}$, or $\xi_{0+}(y)=0$ and $\xi_{0-}(y)\neq 0$ if the integral 
$I_{odd+}$ is infinite, making the equation (\ref{Ieven}) in the form $I_{even}=I_{odd-}$.


\subsection{Thin Brane}

First we are going to analyze fermions localization for the case of a delta-like brane. Replacing, $A(y)=-k_p|y|$, $F(\phi,\pi)=H(\phi,\pi)=0$, $B(y)=0$ in the 
equations (\ref{f(y)}) and (\ref{xi(y)}) we find $f_0(y)=$constant and $\xi_0(y)=\xi_0$. For this case the localization integral for odd dimension is 
\begin{eqnarray}
 I_{odd+}=I_{odd-}=I\equiv |C|^2\int^{+\infty}_{-\infty}dy{}e^{k_p|y|}\rightarrow \infty.
 \label{Int}
\end{eqnarray}
Therefore free Dirac fermions are not localized in both odd and even dimensions. It is the same result obtained by \cite{Bajc:1999mh} for $D=5$. The localization is 
only successful for negative tension brane ($k_p<0$). In order to localize 1/2-spin fermions in the RSM-II framework, we will use, in the next sections, the method
of localization propose by \cite{Kehagias:2000au}, which consists of the introduction of an interaction of fermions with scalar and dilaton fields.

\subsection{Thick Brane without Dilaton Coupling}

When we consider smooth Randall-Sundrum model without dilaton coupling, $A(y)$ is given by (\ref{A}) and can be written as
\begin{eqnarray}
 e^{A(y)}=\frac{\exp{\left[\frac{\beta_p}{2}\tanh^2(by)\right]}}{\left[\cosh(by)\right]^{2\beta_p}}.
\end{eqnarray}
Considering a Yukawa interaction given by $F(\phi,\pi)=\eta_1\phi$, we get
\begin{eqnarray}
 I_{\pm}=|C_\pm|^2\int^{+\infty}_{-\infty}dy[\cosh(by)]^{2(\beta_p\pm a\eta_1/b)}\exp{\left[\frac{\beta_p}{2}\tanh^2{(by)}\right]}. 
 \label{Ipm}
\end{eqnarray}
The evaluation of localization integral is done asymptotically. For $y\rightarrow\pm\infty$, the integrand behavior is proportional to $e^{2b|y|(\beta_p\pm a\eta_1/b)}$ and
the integral (\ref{Ipm}) will converge if 
\begin{eqnarray}
 |\eta_1|>\frac{b\beta_p}{a}.
\end{eqnarray}
Therefore, for odd dimensions, Weyl left fermions $\psi_{0-}(x)$ are localizable if the condition $\eta_1>0$ is 
satisfied. However, if $\eta_1<0$ only Weyl right fermions $\psi_{0+}(x)$ are localizable if the same condition on the $\eta_1$ is true, as well the choice . We conclude that, in odd dimensions, Weyl spinors with different chiralities
are not localizable simultaneously. For even dimensions not exist Weyl spinors on the $p$-brane, thus the signals $\pm$ are related to complex functions $\xi_{0\pm}(y)$.
Therefore according to Sect. \ref{Representation} and Sect. \ref{General case} Dirac's fermions are localizable, for $\eta_1>0$ if $\xi_{0+}(y)=0$ and $\xi_{0-}(y)\neq0$
and for $\eta_1<0$ if $\xi_{0+}(y)\neq0$ and $\xi_{0-}(y)=0$.  

\subsection{Thick Brane with Dilaton Coupling}

    In the case of dilaton coupling with Yukawa interaction i.e., $A(y)$ given by (\ref{A}), $B(y)$ given by (\ref{B}), $F(\phi,\pi)=\eta_2\phi e^{-\lambda\pi}$, and 
$H(\phi,\pi)=0$, the integral (\ref{Ioddpm}) becomes
\begin{eqnarray}
 I_{odd\pm}=|C_\pm|^2\int^{+\infty}_{-\infty}dy\exp{\left[(r-1)A(y)\pm2a\eta_2\int_y dy'e^{\left(r+2\lambda\sqrt{rpM^p}\right)A(y')}\tanh(by')\right]}.
 \label{Id}
\end{eqnarray}
Here we define $\alpha_p\equiv r+2\lambda\sqrt{rpM^p}$, and there are two cases to be analyzed: (i) $\alpha_p=0$ and (ii) $\alpha_p\neq0$. 

\subsubsection{\textit{Case 1:} $\alpha_p=0$}

    In this case, the localization integral (\ref{Id}) can be written as
\begin{eqnarray}
I_{odd\pm} = |C_\pm|^2\int^{+\infty}_{-\infty}dye^{\left[\frac{\beta_p(r-1)}{2}\tanh^2(by)\right]}[\cosh(by)]^{2\left[\beta_p(1-r)\pm a\eta_2/b\right]}.
\label{Ialpha0}
\end{eqnarray}    
The evaluation of localization integrals (\ref{Ialpha0}) is done asymptotically. When $y\rightarrow\pm\infty$, we have
\begin{eqnarray}
I_{odd\pm} \varpropto \left\{ \begin{array}{ll}
         \displaystyle \int^{+\infty}_{-\infty}dy\exp{\left(\pm\frac{2a\eta_2}{b}|y|\right)},                    & \mbox{for $r=1$};\\
         \displaystyle \int^{+\infty}_{-\infty}dy\exp{\left\{2\left[\beta_pb(1-r)\pm a\eta_2\right]|y|\right\}}, & \mbox{for $r\neq1$}.
                     \end{array} \right.
\label{Ialpha1}
\end{eqnarray}
Thus, similarly to the thick brane case, i.e., for $r=1$, we can localized left ($\eta_2>0$) or right ($\eta_2<0$) fermions in odd dimensions. In even dimensions Dirac fermions are
localizable if we choose $\xi_{0+}(y)=0$, $\xi_{0-}(y)\neq0$ and for $\eta_2>0$ or if we choose $\xi_{0+}(y)\neq0$ and $\xi_{0-}(y)=0$ for $\eta_2>0$. However, for $r\neq 1$, the 
analysis is more delicate, because of the the energy density (\ref{T00}) must be finite. In this case, the localization will be successful if $\beta_pb(1-r)\pm a\eta_2<0$. 

    For $0<r<1$, this condition can be written as $\displaystyle |\eta_2|>\beta_pb(1-r)/a$ and we conclude that, in odd dimensions, if $\eta_2>0$ left handed fermions are localizable, 
but if $\eta_2<0$ only right handed fermions are localizable. For even dimensions, the Dirac spinor $\psi_0(x)$ will be localized for $\eta_2>0$ if we choose $\xi_{0+}(y)=0$,
$\xi_{0-}(y)\neq0$ and for $\eta_2<0$ if we choose $\xi_{0+}(y)\neq0$ and $\xi_{0-}(y)=0$. 

    For $r=(p+1)/2>1$, the localization condition can be written as $|\eta_2|<\beta_pb(D-3)/2a$, where $D$ is the bulk dimensionality. Therefore, both $\eta_2>0$ and $\eta_2<0$, the
Dirac spinor $\psi_0(x)$ is localized for both odd and even dimensions. Beyond this, in this case, the localization mechanism is improved with the increase of bulk dimensionality.

\subsubsection{\textit{Case 2:} $\alpha_p\neq0$}

    In this case we define
\begin{eqnarray}
I(y)\equiv\int_y dy'e^{\alpha_pA(y')}\tanh(by'),
\end{eqnarray}
and for $y\rightarrow\pm\infty$ we have
\begin{eqnarray}
 I(y)\varpropto -\frac{1}{2\alpha_p}e^{-\alpha_p|y|},
\end{eqnarray}
and therefore the localization integral (\ref{Id}) becomes
\begin{eqnarray}
 I_{odd\pm} \varpropto \left\{ \begin{array}{ll}
         \displaystyle \int^{+\infty}_{-\infty}dy\exp{\left(\mp\frac{a\eta_2}{\alpha_p}e^{-\alpha_p|y|}\right)},          & \mbox{for $r=1$};\\
         \displaystyle \int^{+\infty}_{-\infty}dy\exp{\left[-(r-1)\beta_pb|y|\mp\frac{a\eta_2}{\alpha_p}e^{-\alpha_p|y|}\right]}, & \mbox{for $r\neq1$}.
                     \end{array} \right.
\label{Ialpha2}
\end{eqnarray}
    For $r=1$, we will analyze two situations: $\alpha_p<0$ and $\alpha_p>0$. When $\alpha_p<0$, according (\ref{Ialpha2}), for odd dimensions right handed fermions $\psi_{0+}(x)$ 
are localized if $\eta_2>0$ or left handed fermions are localized if $\eta_2<0$, but Weyl spinors with different chiralities are not localizable simultaneously. In even 
dimensions the Dirac spinor $\psi_0(x)$ will be localized if we choose $\xi_{0+}(y)\neq0$, $\xi_{0-}(y)=0$ for $\eta_2>0$ or we choose $\xi_{0+}(y)=0$, $\xi_{0-}(y)\neq0$ for 
$\eta_2<0$. When $\alpha_p>0$, the first integral of the equation (\ref{Ialpha2}) is always divergent, and therefore can not possible localize spinors both in odd as even dimensions.

    For $0<r<1$, we will also analyze $\alpha_p<0$ and $\alpha_p>0$. When $\alpha_p<0$, we have the localization integral $I_{odd\pm}$ asymptotically identical to case $r=1$,
providing the same results. When $\alpha_p>0$, the integral $I_{odd\pm}$ becomes
\begin{eqnarray}
 I_{odd\pm}=\int^{+\infty}_{-\infty}dye^{(1-r)\beta_pb|y|}\rightarrow\infty.
\end{eqnarray}
Thus, can not possible localize spinors both in odd as even dimensions.

    For $r=(p+1)/2>1$, we obtained the same results from the case $r=1$ if $\alpha_p<0$. However, for
$\alpha_p>0$, we have
\begin{eqnarray}
 I_{odd\pm}\varpropto \int^{+\infty}_{-\infty}dye^{-\frac{(D-3)}{2}\beta_pb|y|}<\infty,
\end{eqnarray}
and we conclude that the Dirac spinor $\psi_0(x)$ is localized for both odd and even dimensions, independent of the parameter $\eta_2$.

\subsubsection{Dilaton Derivate Coupling}	

    Finally, we will analyze the localization for $F(\phi,\pi)=0$ and $H(\phi,\pi)=h\pi(y)$, where $h$ is a real coupling constant. In this case, the localization integral 
(\ref{Ioddpm}) becomes
\begin{eqnarray}
 I_{odd\pm}=|C_\pm|^2\int^{+\infty}_{-\infty}dy\exp{\left[(r-1)A(y)\mp4h\sqrt{\frac{pM^p}{r}}e^{rA(y)}\right]}.
\end{eqnarray}
To verify the convergence of this integral, we must again evaluate it asymptotically,
\begin{eqnarray}
I_{odd\pm} \varpropto \left\{ \begin{array}{ll}
         \displaystyle \int^{+\infty}_{-\infty}dy\exp{\left(\mp4h\sqrt{pM^p}e^{-2b\beta_p|y|}\right)},                    & \mbox{for $r=1$};\\
         \displaystyle \int^{+\infty}_{-\infty}dy\exp{\left[-2b\beta_p(r-1)|y|\mp4h\sqrt{\frac{pM^p}{r}}e^{-2b\beta_pr|y|}\right]}, & \mbox{for $r\neq1$}.
                     \end{array} \right.
\label{Ialpha3}
\end{eqnarray}

    For $r=1$, the integral $I_{odd\pm}$ diverge, because $e^{-2b\beta_p|y|}\rightarrow 0$ when $y\rightarrow\pm\infty$, and therefore $\mp4h\sqrt{pM^p}e^{-2b\beta_p|y|}\rightarrow$
constant. Thus, we conclude that spinorial fields are not localizable in odd and even dimensions.

    For $0<r<1$, the integral $I_{odd\pm}$ becomes
\begin{eqnarray}
  I_{odd\pm}\varpropto\int^{+\infty}_{-\infty}dye^{2b\beta_p(1-r)|y|}\rightarrow\infty.
\end{eqnarray}
Therefore we have that Dirac's fermions are not localized in odd or even dimensions.

    For $r=(p+1)/2>1$, we have
\begin{eqnarray}
  I_{odd\pm}\varpropto\int^{+\infty}_{-\infty}dye^{-2b\beta_p\frac{(D-3)}{2}|y|}<\infty,
\end{eqnarray}
showing that Dirac spinor are localizable for both odd and even dimensions.

\section{Resonances}
\label{Resonances}

    Now let us turn our attention to the massive modes of Dirac equation. We are interested only in the study of resonant modes and must use the
  transfer matrix method.  They depend strongly of the 
form of the potential and, therefore, of the model considered. Here we will study the model similar to that found in \cite{Kehagias:2000au}.

\subsection{Review of the Transfer Matrix Method}

    Here we will give the details of the program used to compute the transmission coefficients by transfer matrix \cite{Landim:2011ki, Landim:2011ts, 
Alencar:2012en}. First, we write the equation of motion for massive modes as a Schr\"odinger like equation, as demonstrated in (\ref{Schoedinger equation}). Due to 
similarity with usual quantum mechanics, we calculate the transmission coefficient, but numerically. For this, we will use the transfer matrix method. This method is 
based on the concept of potential barrier. The solutions before and after the potential barrier are plane waves, and using this fact it is possible to approximate 
this type of solution in each barrier. 

\begin{figure}[!htb]
\centering
\includegraphics[width=0.5\textwidth]{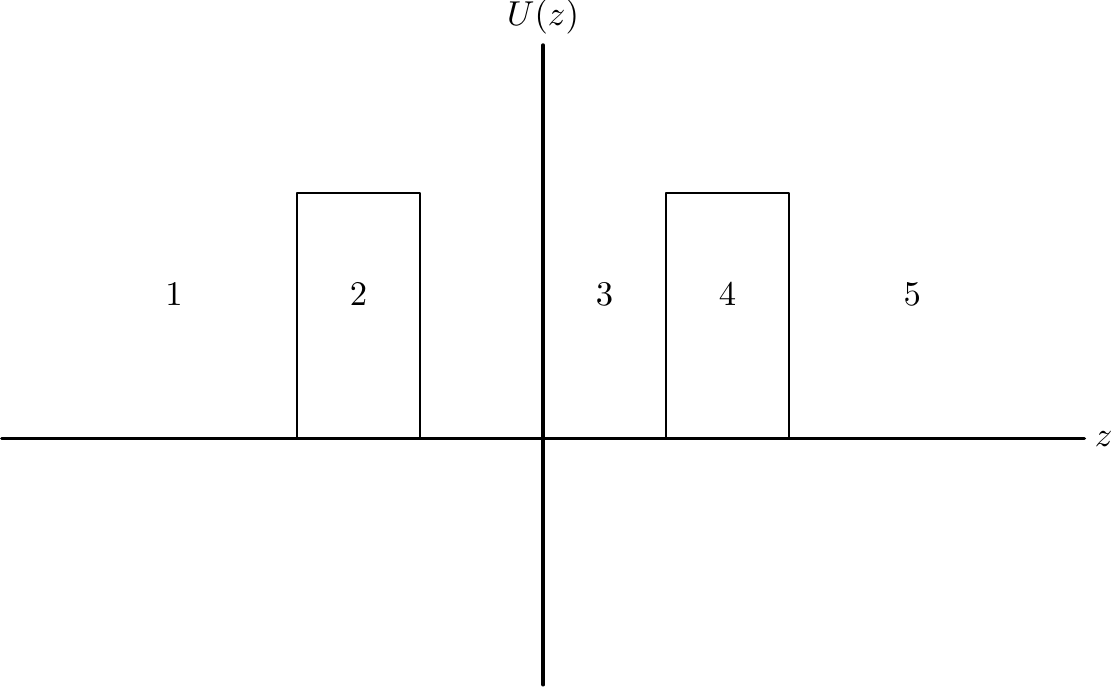}
\caption{The double potential barrier.}
\label{Double}
\end{figure}

    We consider as an example the simple case given by the double well barrier as shown in Fig. \ref{Double}. The solution of the Schr\"odinger equation for each 
region is given by
\begin{eqnarray}
 \bar{\Phi}_j(z)=A_je^{ik_jz}+B_je^{ik_jz}, \quad k_j\equiv \sqrt{2(E-U_j)}, \quad j=1,2,3,4,5.
\end{eqnarray}

\begin{figure}[!htb]
\centering
\includegraphics[width=0.5\textwidth]{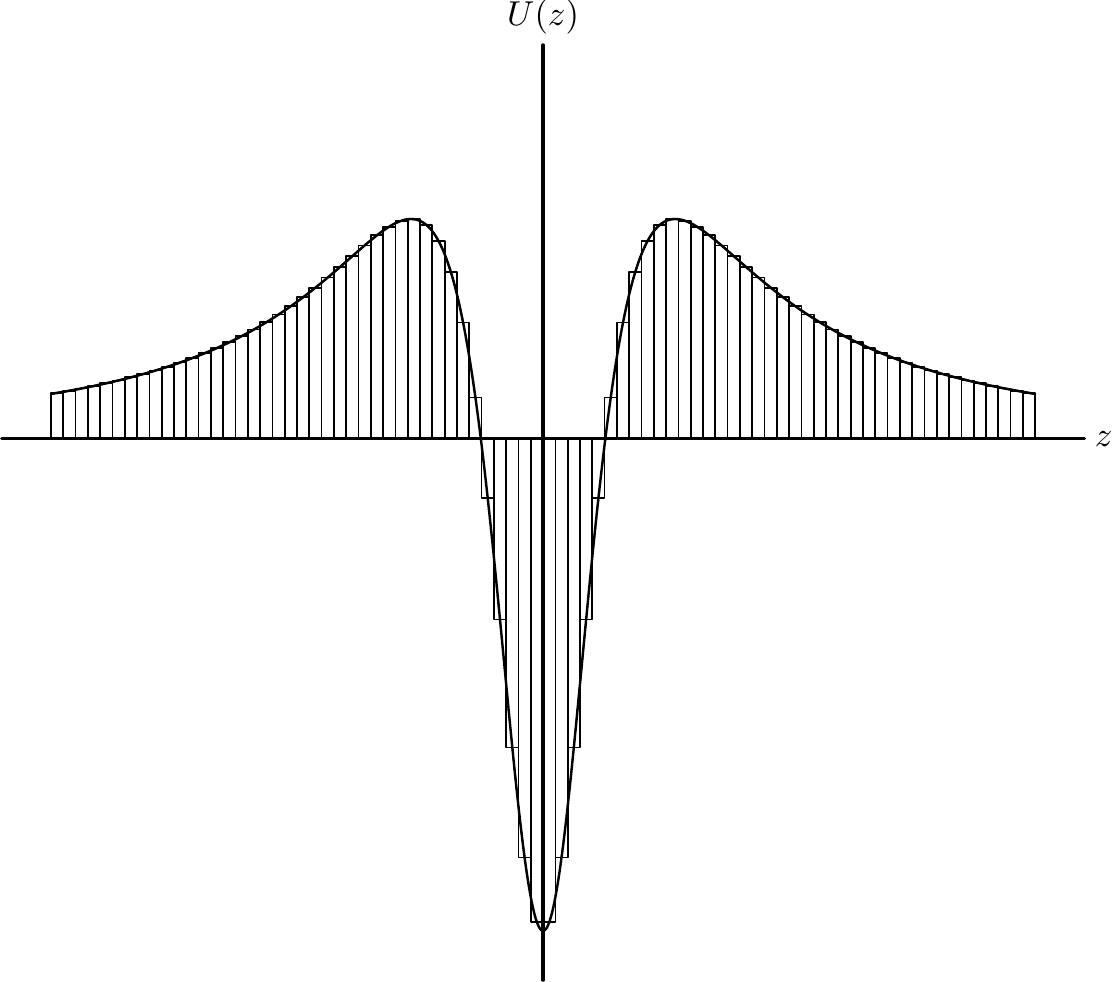}
\caption{Potential approximation by barrier}
\label{PotentialVolcano}
\end{figure}

The idea therefore is to apply the results above to a more general case. For this purpose the potential can be approximated by a series of barriers showed in the 
Fig. \ref{PotentialVolcano}. The Schr\"odinger equation must be solved for each interval $z_{j-1}<z<z_j$ , where we have approximate the potential by
\begin{eqnarray}
 U(z)=U(\bar{z}_{j-1})=U_{j-1}, \quad \bar{z}_{j-1}=(z_j+z_{j-1})/2.
\end{eqnarray}
The solution in this interval is
\begin{eqnarray}
 \bar{\Phi}_{j-1}(z)=A_{j-1}e^{ik_{j-1}z}+B_{j-1}e^{ik_{j-1}z}, \quad k_{j-1}\equiv \sqrt{m^2-U_{j-1}}.
\end{eqnarray}
the continuity of the $\bar{\Phi}_{j-1}(z)$ and $\bar{\Phi}'_{j-1}(z)$ at $z=z_j$ give us
\begin{eqnarray}
 \left(\begin{array}{c}
 A_j\\
 B_j\\
\end{array}\right)=M_j\left(\begin{array}{c}
                         A_{j-1}  \\
                         B_{j-1} \\
                        \end{array}\right).
\end{eqnarray}
In the above equation we have that
\begin{eqnarray}
 M_j=\frac{1}{2k_j}\left[\begin{array}{cc}
                         (k_j+k_{j-1})e^{-i(k_j-k_{j-1})z_j} & (k_j-k_{j-1})e^{-i(k_j+k_{j-1})z_j}\\
                         (k_j-k_{j-1})e^{i(k_j+k_{j-1})z_j} & (k_j+k_{j-1})e^{i(k_j-k_{j-1})z_j}\\
                         \end{array}\right],
\end{eqnarray}
and performing this procedure iteratively we reach
\begin{eqnarray}
 \left(\begin{array}{c}
 A_N\\
 B_N\\
\end{array}\right)=M\left(\begin{array}{c}
                         A_0  \\
                         B_0 \\
                        \end{array}\right),
\end{eqnarray}
where,
\begin{equation}
 M = M_N M_{N-1}...M_2 M_1.
\end{equation}
The transmission coefficient is therefore given by
\begin{eqnarray}
 T=\frac{1}{|M_{22}|^2}.
\end{eqnarray}
This expression must be a function of $m^2$. To obtain the numerical resonance values we choose the $z_{max}$ to satisfy $U(z_{max})\sim 10^{-4}$ and $m^2$ runs from
$U_{min} = U(z_{max})$ to $U_{max}$ (the maximum potential value). We divide $2z_{max}$ by $10^4$ or $10^5$ such that we have $10^4+1$ or $10^5+1$ transfer matrices.

\subsection{The Case $F(\phi,\pi)=\eta_2\phi e^{-\lambda\pi}$ and $H(\phi,\pi)=0$}

   In this subsection we will analyze the resonances for the potential $F(\phi,\pi)=\eta_2\phi e^{-\lambda\pi}$ and $H(\phi,\pi)=0$, 
with $r=1$, $\eta_2=1$ and for the dimensionalities $D=5,10$, for $s=1$, $s=3$ and $s=5$  for two cases, $\lambda=0,4$ and $\lambda=1/\sqrt{pM^p}$.

\begin{figure}[h]

\center
\subfigure{\includegraphics[width=0.45\textwidth]{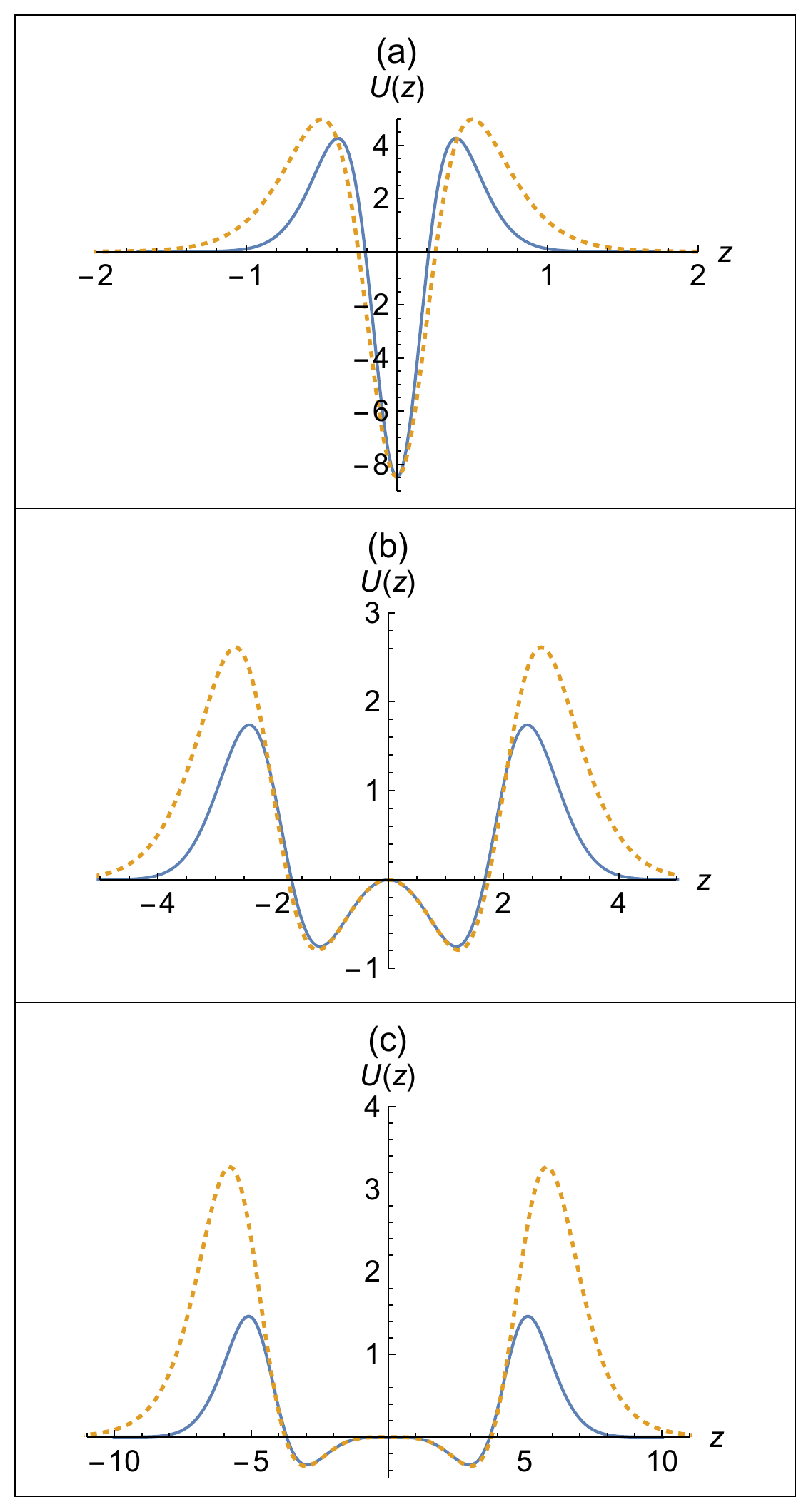}}
\quad
\subfigure{\includegraphics[width=0.45\textwidth]{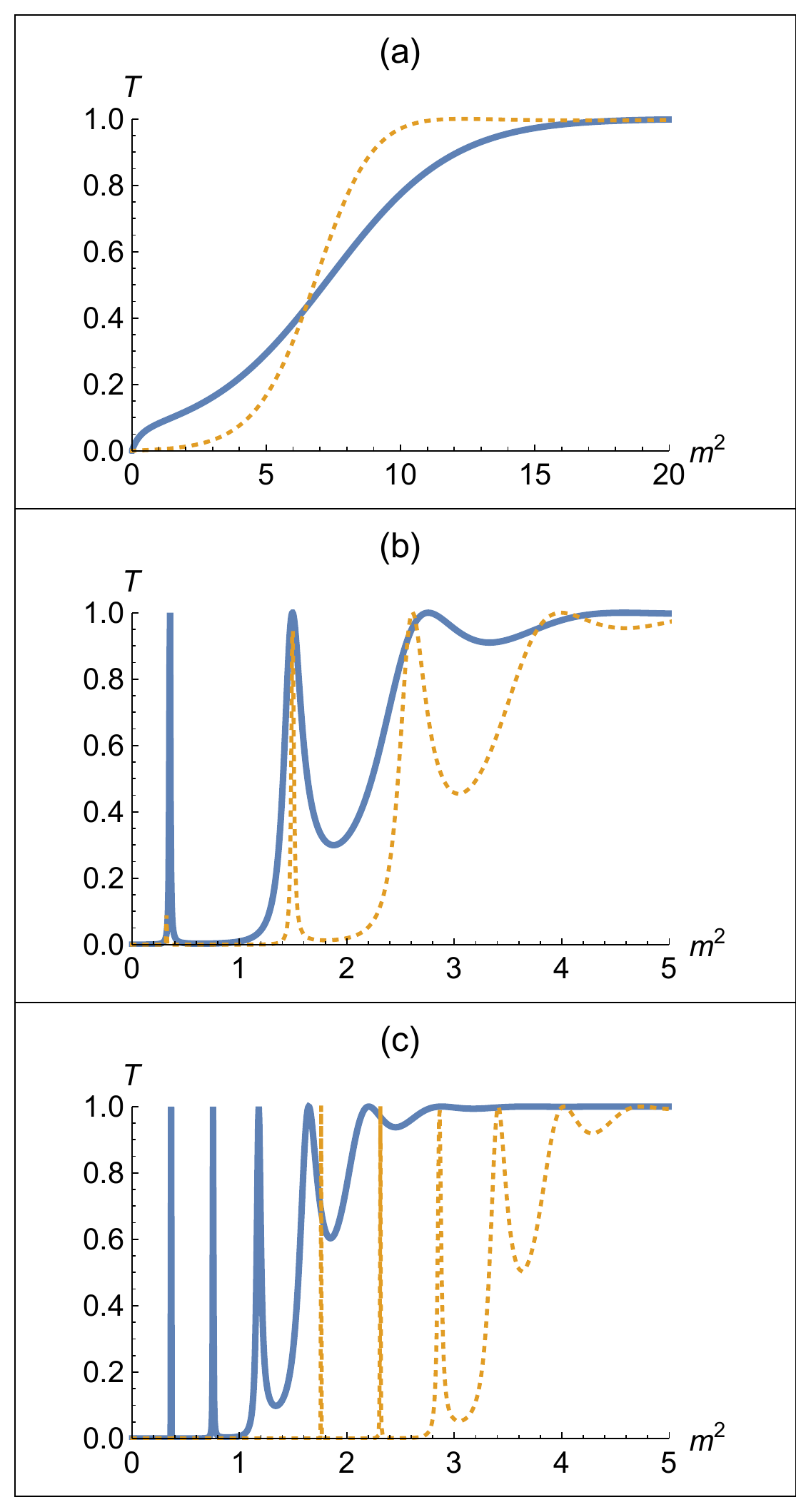}}
\caption{Potentials $U(z)$ and Transmission coefficient $T(m^2)$ for massive left fermions (odd dimensions) and Dirac fermions (even dimensions choosing $\xi_{n-}(x)$)
with $\lambda=0,4$ for $D=5$ (line) and $D=10$ (dotted), with $\eta=1$ and $s=1$ (a, d), $s=3$ (b, e) and $s=5$ (c, f).}
\label{Potentials2}
\end{figure}

    In the case where $\lambda=0,4$, we will analyze left and right handed resonance fermions (for odd dimensions) and Dirac spinor (even dimensions). We show in Fig.
\ref{Potentials2} the plots for the potentials and of the transmission coefficient for left fermions (odd dimensions) and Dirac spinor (even dimensions, using the function 
$\xi_{n-}(y)$). Particularly, for even dimensions, if the transmission coefficients of $\xi_{n+}(x)$ or $\xi_{n-}(x)$ show peaks for $m^2\leq U_{max}$, it is 
sufficient for that to be found resonance modes in Dirac's fermions. This fact is a consequence of decomposition \ref{SolPar}, that is different of odd case 
\ref{decomposition}. 

    The potential for left handed and Dirac fermions for $s=1$ is very similar to a double barrier and we should expect the existence of resonant modes 
\cite{Landim:2011ki, Landim:2011ts, Alencar:2012en}. But for right handed fermions, when $s=1$ this will not be true, because the potential in this case is a single 
barrier. We observe that, in this case, the deformed topological defects are responsible for the appearance of resonance peaks. The number of resonance peaks is 
very similar to the previous case but not their positions. The table \ref{Tablecase2a} shows the values of the peaks of resonance. We observed that, for $D=5,10$ 
and $s=1$, although the potential takes the form of a double barrier, no resonance peaks were found.

\begin{table}[h]
\caption{Peaks of resonances for massive left handed fermions (odd dimension) or Dirac fermions (even dimensions choosing $\xi_{n-}(x)$) for $\lambda=0,4$.}
\label{Tablecase2a}
\begin{center}
 \begin{tabular}{|c|c|c|c|c|c|c|}
  \hline
   Topological Defects   & \multicolumn{6}{|c|}{Resonances Peaks ($m^2$)}  \\
  \hline
                   $s=1$ &    -       &    -   &   -   &     -     &    -     &  -  \\
  \hline
                   $s=3$ &    0,35    &    -   &   -   &    0,32   &    1,5   &  -  \\
\hline
                   $s=5$ &    0,36    &  0,75  &   -   &    1,75   &   2,3    &  2,85 \\
 \hline
 Dimensionality          & \multicolumn{3}{|c|}{$D=5$} & \multicolumn{3}{|c|}{$D=10$}\\
 \hline
 \end{tabular}
 \end{center}
\end{table}

\begin{figure}[h]

\center
\subfigure{\includegraphics[width=0.45\textwidth]{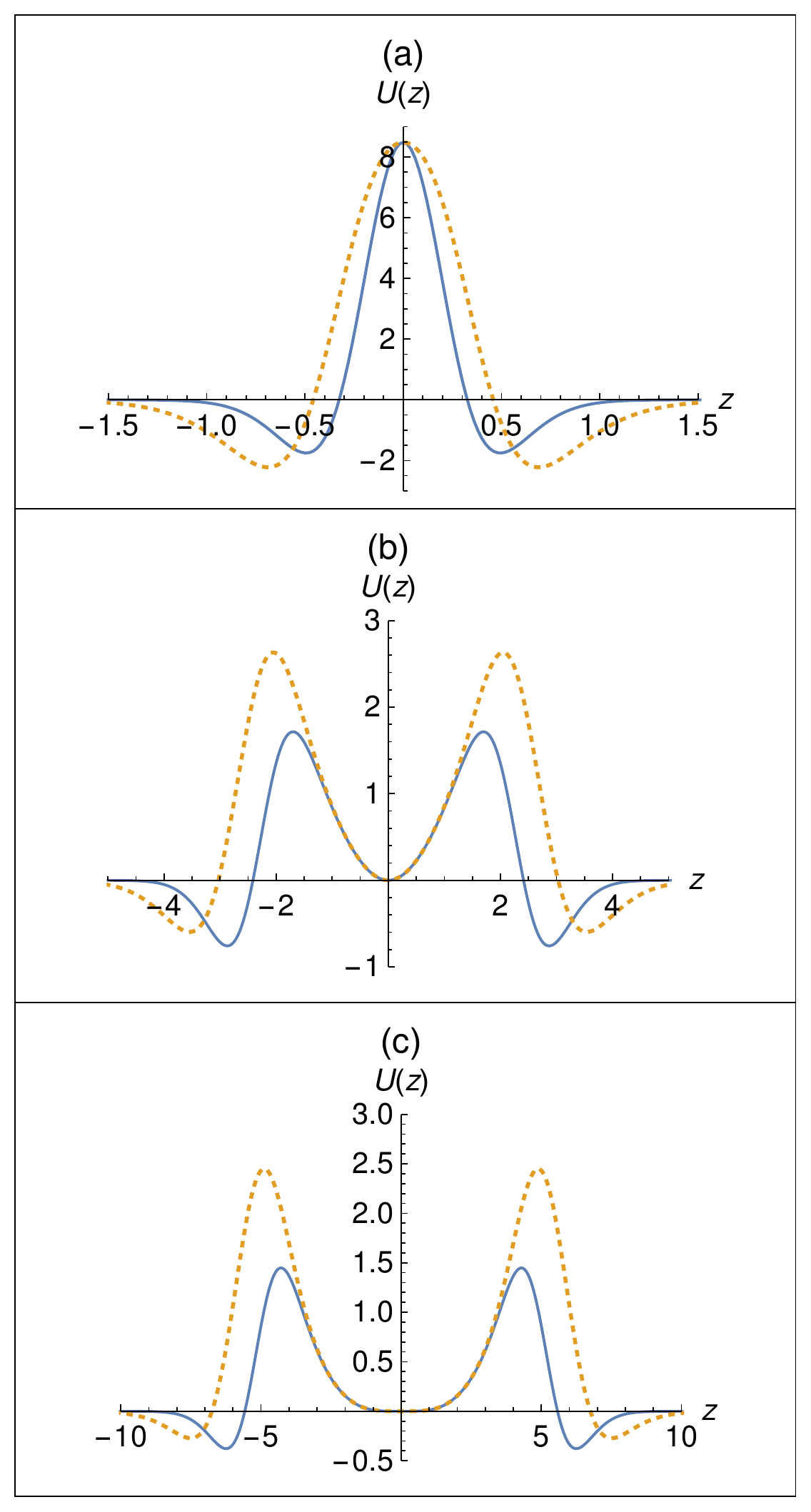}} 
\qquad
\subfigure{\includegraphics[width=0.45\textwidth]{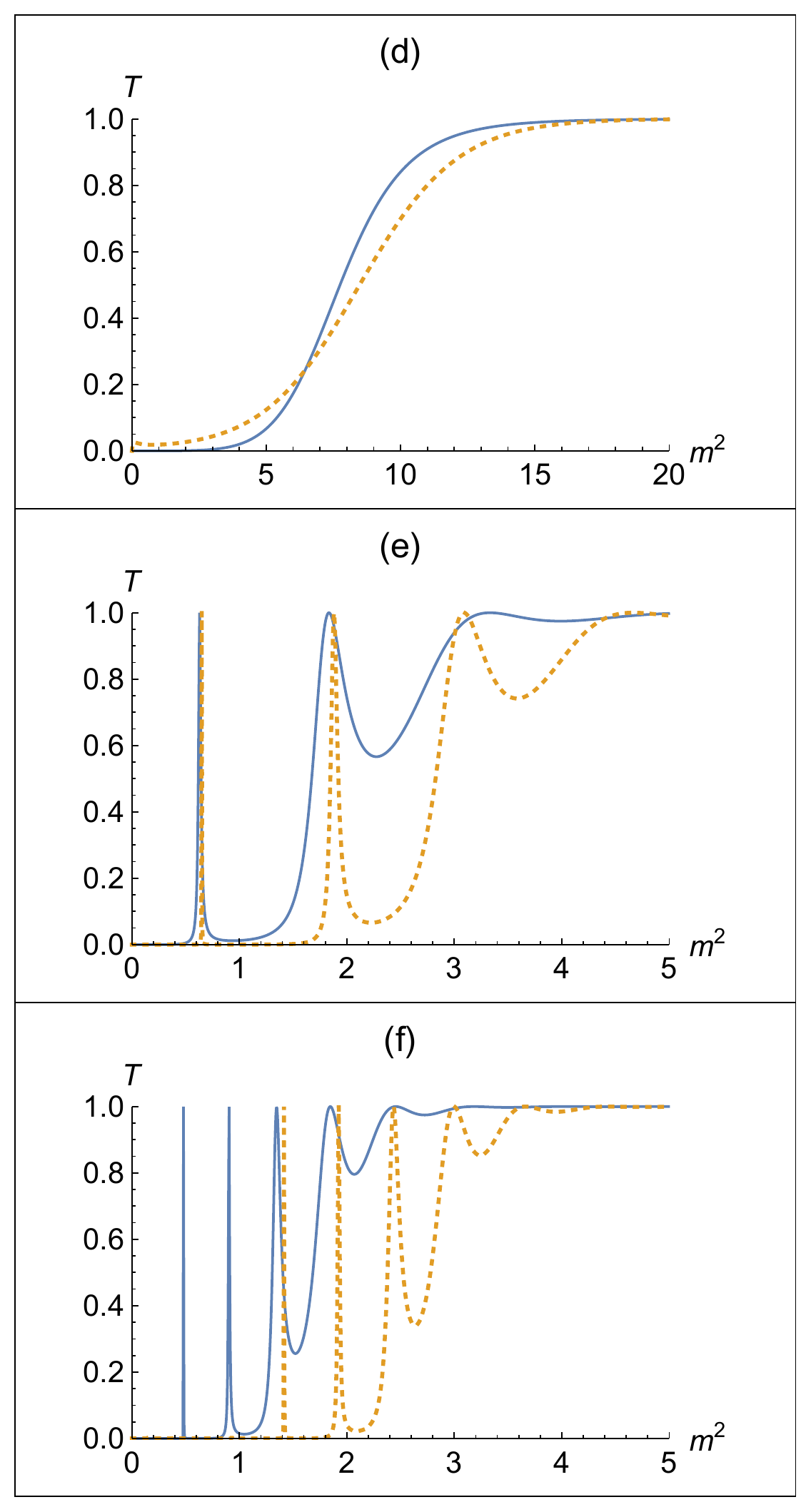}}
\caption{Potentials $U(z)$ and Transmission coefficient $T(m^2)$ for massive right fermions (odd dimensions) and Dirac fermions (even dimensions choosing $\xi_{n+}(x)$)
with $\lambda=0,4$ for $D=5$ (line) and $D=10$ (dotted), with $\eta=1$, $M=1$ and $s=1$ (a, d), $s=3$ (b, e) and $s=5$ (c, f).}
\label{Potentials3}
\end{figure}

    For right handed fermions (odd dimensions) and Dirac fermions (even dimensions, using the function $\xi_{n_+}(y)$), the potentials and transmission coefficient 
are showed in Fig. \ref{Potentials3}. We observed that, in this case, the potentials do not have the form of a double barrier and therefore there are no resonant 
modes. For this reason we do not find peaks of probability for massive fermions coupled with the dilaton when $\lambda=0,4$, as shown in Fig. \ref{Potentials3} (d).
The results obtained enforces the previous affirmation that the resonances are strongly dependent on the topological defect parameter $s$. The table \ref{Tablecase2b}
shows the peaks of resonance.

\begin{table}[h]
\caption{Peaks of resonances for massive right handed fermions.}
\label{Tablecase2b}
\begin{center}
 \begin{tabular}{|c|c|c|c|c|}
  \hline
   Topological Defects    & \multicolumn{4}{|c|}{Resonances Peaks ($m^2$)}         \\
  \hline
                    $s=1$ &  -            &      -    &      -     &     -         \\
  \hline
                    $s=3$ &      0,63     &      -    &    0,65    &     1,9       \\
\hline
                    $s=5$ &      0,48     &    0,9    &    1,4    &      1,9       \\
 \hline
 Dimensionality           &\multicolumn{2}{|c|}{$D=5$}&\multicolumn{2}{|c|}{$D=10$}\\
 \hline
 \end{tabular}
 \end{center}
\end{table}

\begin{figure}[h]

\center
\subfigure{\includegraphics[width=0.45\textwidth]{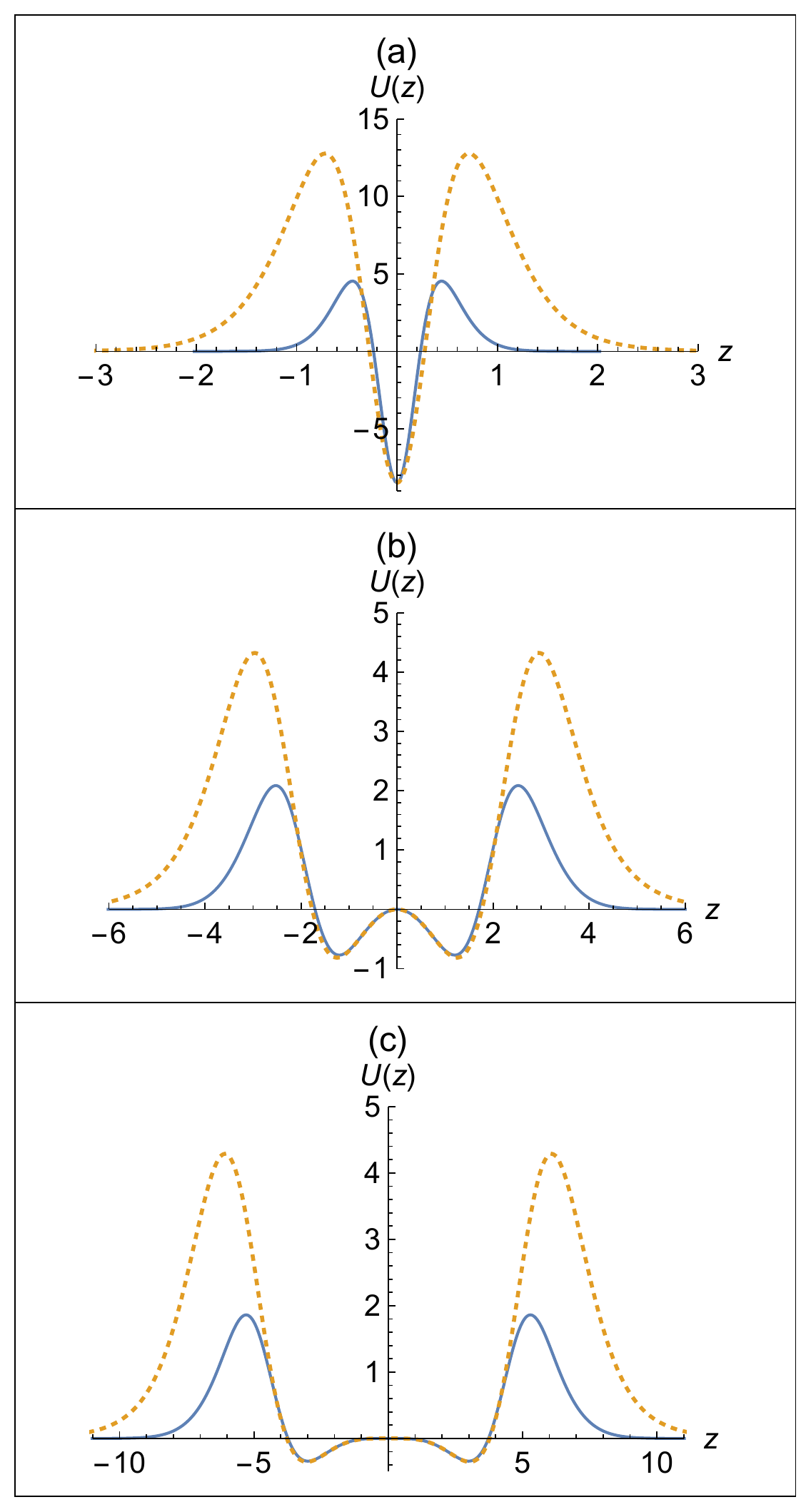}} 
\quad
\subfigure{\includegraphics[width=0.45\textwidth]{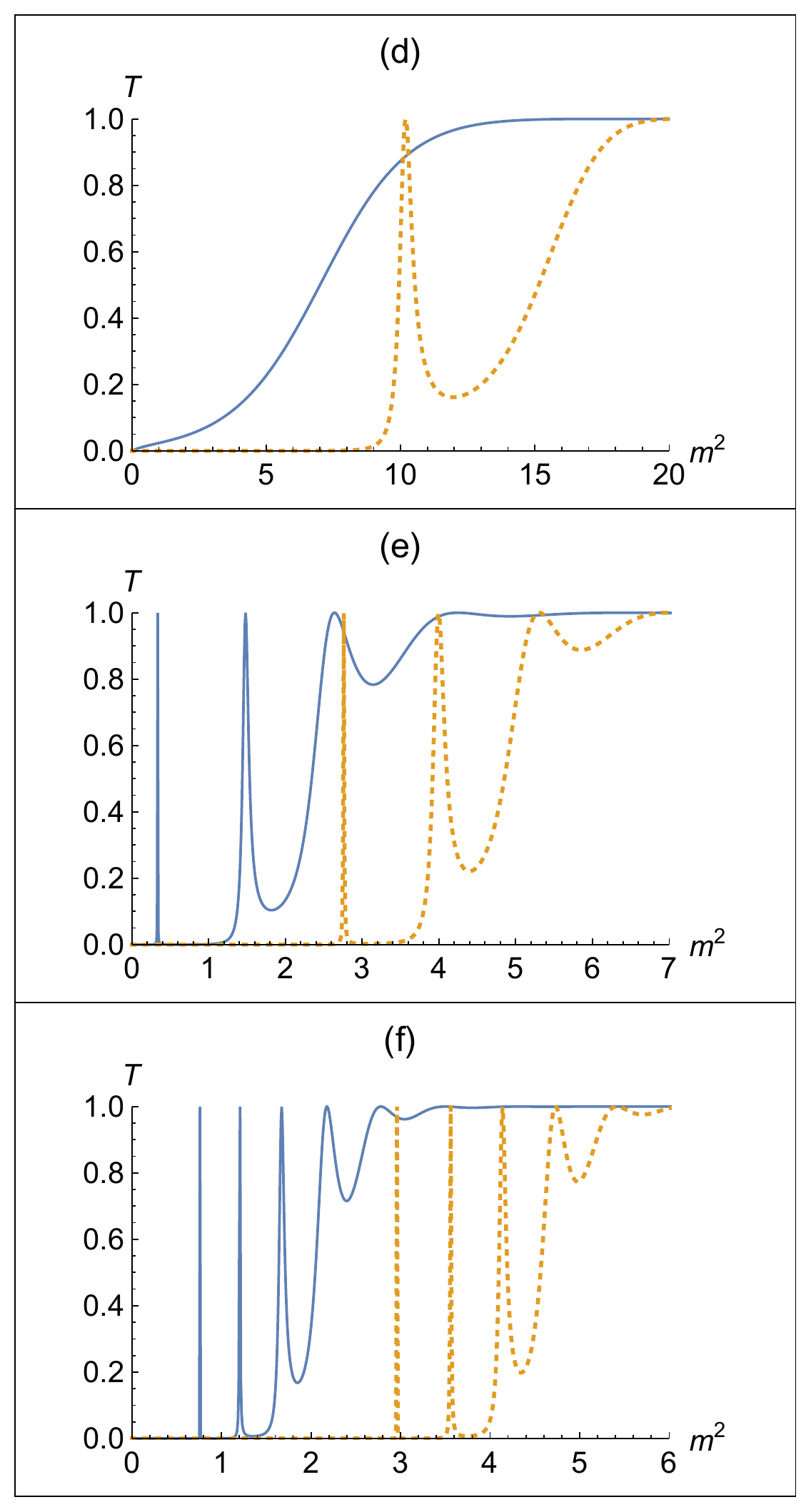}}
\caption{Potentials $U(z)$ and Transmission coefficient $T(m^2)$ for massive left fermions (odd dimensions) and Dirac fermions (even dimensions, choosing 
the function $\xi_{n-}(x)$) with $\lambda=1/(2\sqrt{p})$) for $D=5$ (line) and $D=10$ (dotted), with $\eta=2$ and $s=1$ (a, d), $s=3$ (b, e) and $s=5$ (c, f).}
\label{Potentials1}
\end{figure}

    As said previously, the peaks of resonances are strongly dependent on the $s$. For $D=5$, if the potential looks like the double well 
barrier, we can observe resonances. But, other hand, if the potential is a one single barrier we cannot observe resonances. For this reason we only plot the 
potentials and transmission coefficient for left handed (odd dimensions) and Dirac (even dimensions, using function $\xi_{n-}(y)$) fermions for the case 
$\lambda=1/(2\sqrt{p})$ (Fig. \ref{Potentials1}). However, we discovered in this case that in $D=10$ and $s=1$ we find resonant modes for Dirac fermions, in
even dimensions. In other words, we discovered that the resonance peaks are dependent of the bulk dimensionality. The table \ref{Tablecase1} shows resonance peak
in $m^2\simeq 10,5$, when $D=10$, while for $D=5$ no resonant peaks appear.

\begin{table}[h]
\caption{Peaks of resonances for massive left handed fermions.}
\label{Tablecase1}
\begin{center}
 \begin{tabular}{|c|c|c|c|c|c|c|}
  \hline
   Topological Defects   & \multicolumn{6}{|c|}{Resonances Peaks ($m^2$)}  \\
  \hline
                   $s=1$ &    -       &  -     &  -    &   10,5    &    -     &  -  \\
  \hline
                   $s=3$ &    0,4     &   1,5  &  -    &    2,8    &    4,0   &  -  \\
\hline
                   $s=5$ &    0,8     &   1,2  &  1,7  &     3,0    &   3,6   &  4,2 \\
 \hline
 Dimensionality          & \multicolumn{3}{|c|}{$D=5$} & \multicolumn{3}{|c|}{$D=10$}\\
 \hline
 \end{tabular}
 \end{center}
\end{table}

\subsection{The Case $F(\phi,\pi)=0$ and $H(\phi,\pi)=h\pi$}

    Here we consider the next important case, which is known as coupling with dilaton derivative, $F(\phi,\pi)=0$ and $H(\phi,\pi)=h\pi$. These cases has already been 
studied in \cite{Liu:2013kxz} for $D=5$. But we will generalize it to the case with co-dimension one and in the light of the Transfer Matrix Method.
   
\begin{figure}[h]
\center
\subfigure{\includegraphics[width=0.45\textwidth]{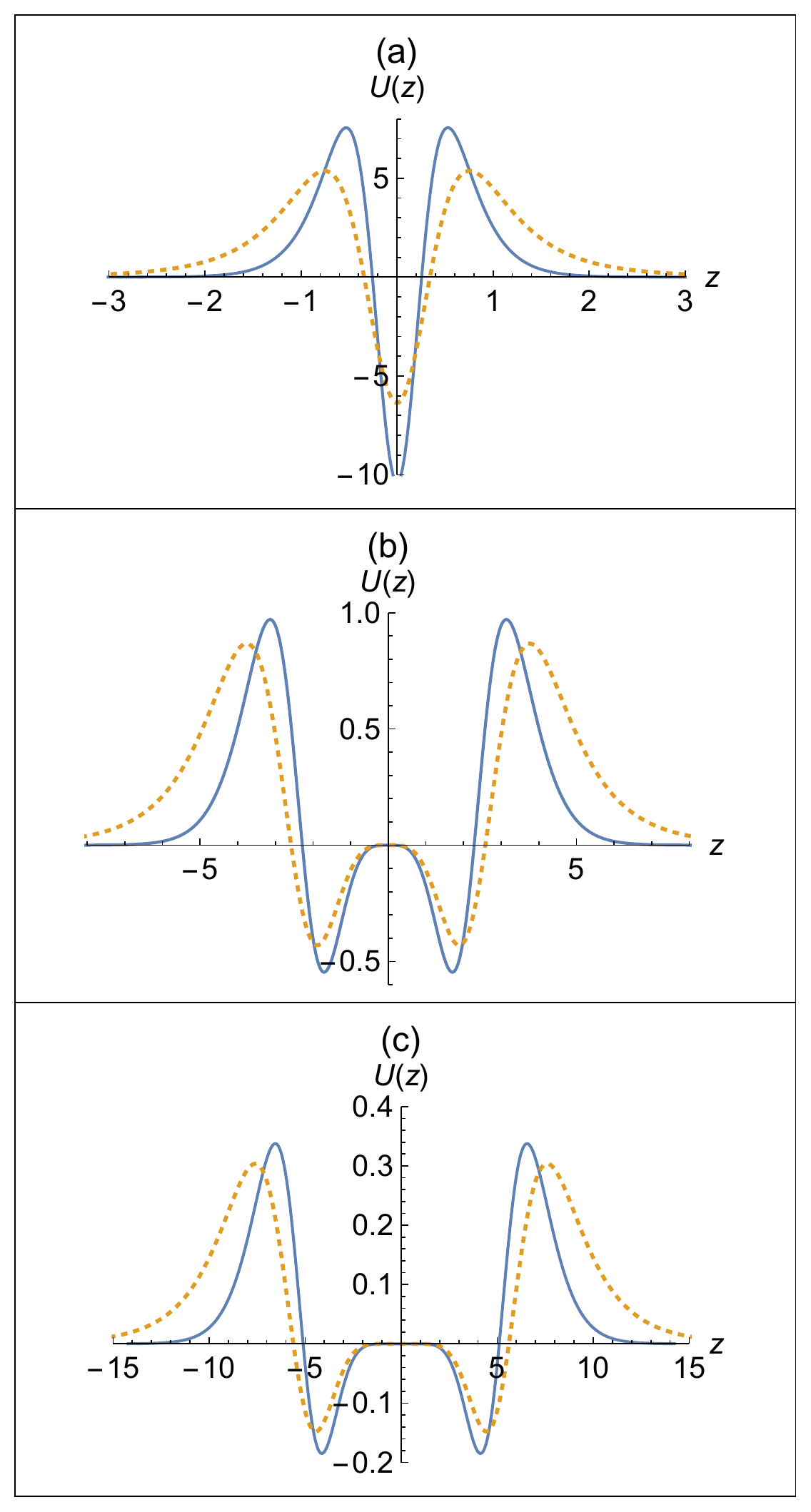}}
\qquad
\subfigure{\includegraphics[width=0.45\textwidth]{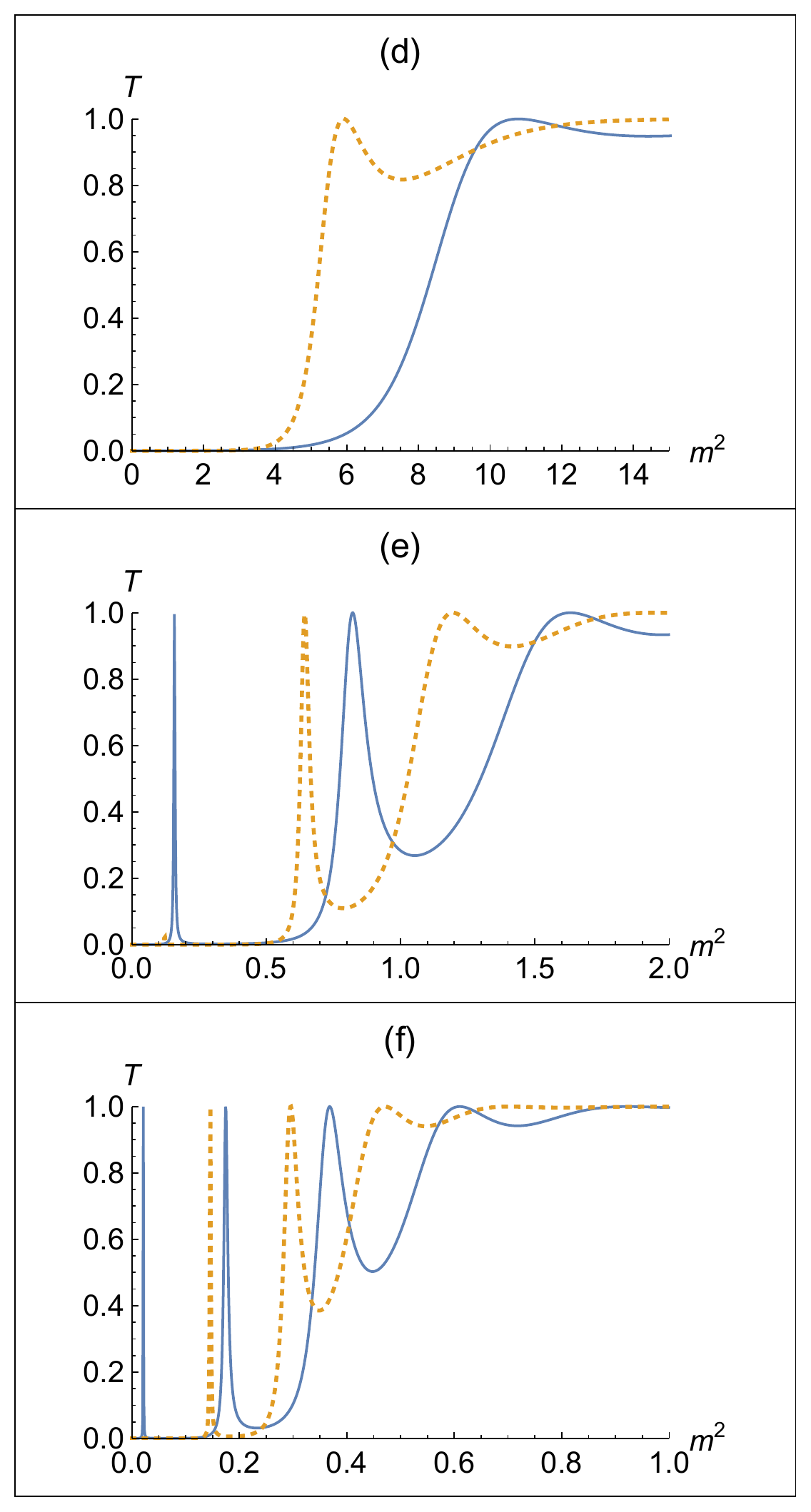}}
\caption{Potentials $U(z)$ and Transmission coefficient $T(m^2)$ for massive left fermions for $D=5$ (line) and $D=10$ (dotted), with $\eta=1$ and 
        $s=1$ (a, d), $s=3$ (b, e) and $s=5$ (c, f) in derivate dilaton setup.}
\label{Potentials5}
\end{figure}

    We show in Fig. \ref{Potentials5} (a) the graphics of the left handed fermion potential with the coupling $H(\phi,\pi)=h\pi$. In Fig. \ref{Potentials5} (d) 
we show the transmission coefficient, when $D=5,10$ and $s=1$. In Figs. \ref{Potentials5} (b), (c) and \ref{Potentials5} (e), (f) we show 
the graphics of the potential and transmission coefficient with $s=3,5$, respectively. Here we note that the increase of the dimensionality leads a decrease in height, which on the other hand is compensated by a gain in the potential width. Also we note that, in this
case the resonance peaks occur at a lower energy scale. For example, in case $F(\phi,\pi)=\eta\phi e^{-\lambda\pi}$ and $H(\phi,\pi)=0$, $s=3$ and $D=5$ (table 
\ref{Tablecase1}), the first peak of resonance occur at $m^2\simeq 0,4$, while for this case (Fig. \ref{Potentials5} (d)) the same occur in $m^2\simeq0,15$.

\begin{table}[h]
\caption{Peaks of resonances for massive right handed fermions.}
\begin{center}
 \begin{tabular}{|c|c|c|c|c|}
  \hline
   Topological Defects    & \multicolumn{4}{|c|}{Resonances Peaks ($m^2$)} \\
  \hline
                           $s=1$ &  -    &  -  &    - &     -     \\
  \hline
                           $s=3$ & 0,16  & -   & 0,65 &     -        \\
\hline
                           $s=5$ & 0,02 & 0,17 & 0,15 &     -        \\
 \hline
 Dimensionality         & \multicolumn{2}{|c|}{$D=5$}  & \multicolumn{2}{|c|}{$D=10$}\\
 \hline
 \end{tabular}
 \end{center}
\label{TableDilaton-1}

\end{table}

\begin{figure}[h]

\center
\subfigure{\includegraphics[width=0.45\textwidth]{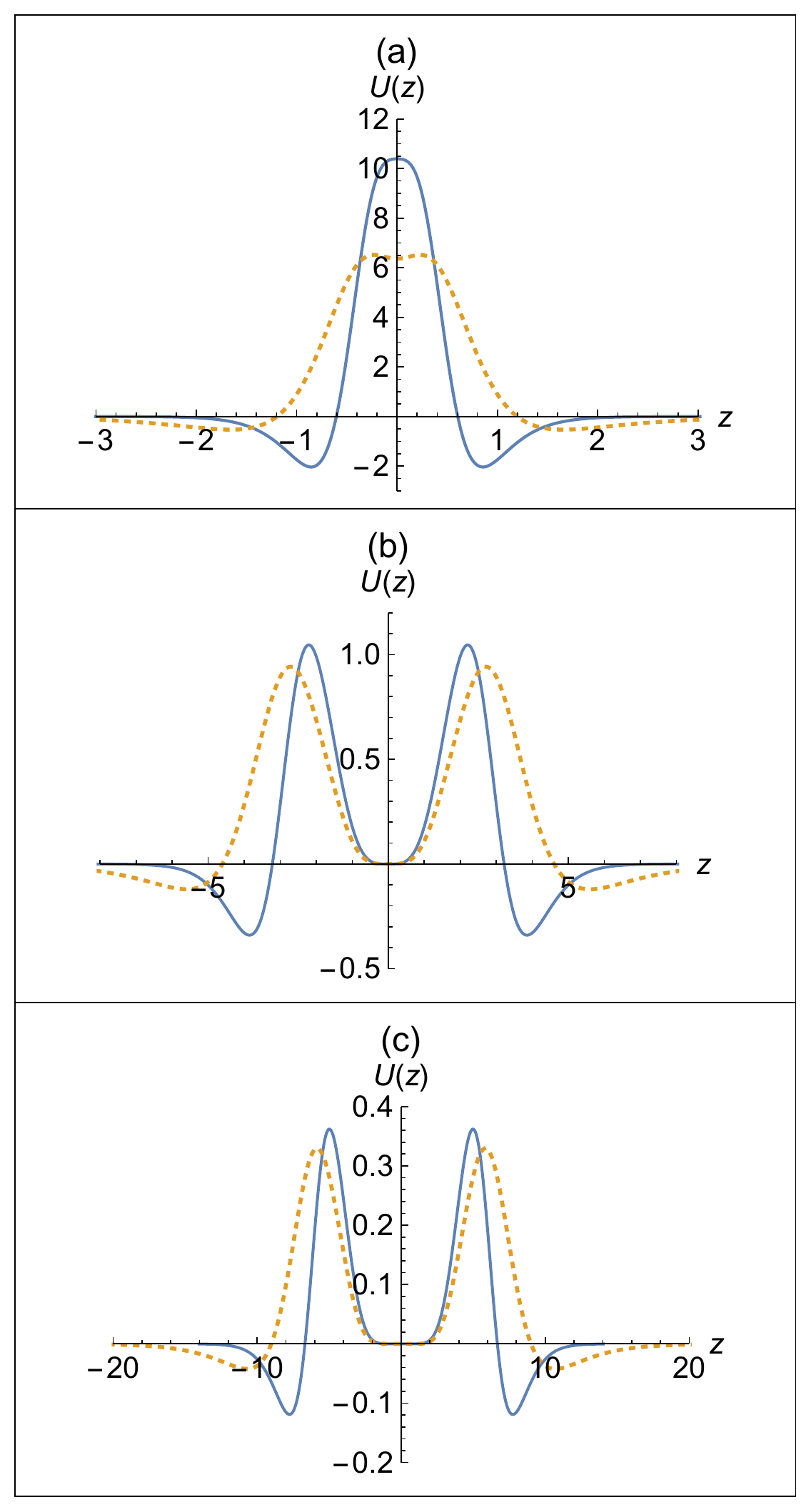}}
\qquad
\subfigure{\includegraphics[width=0.45\textwidth]{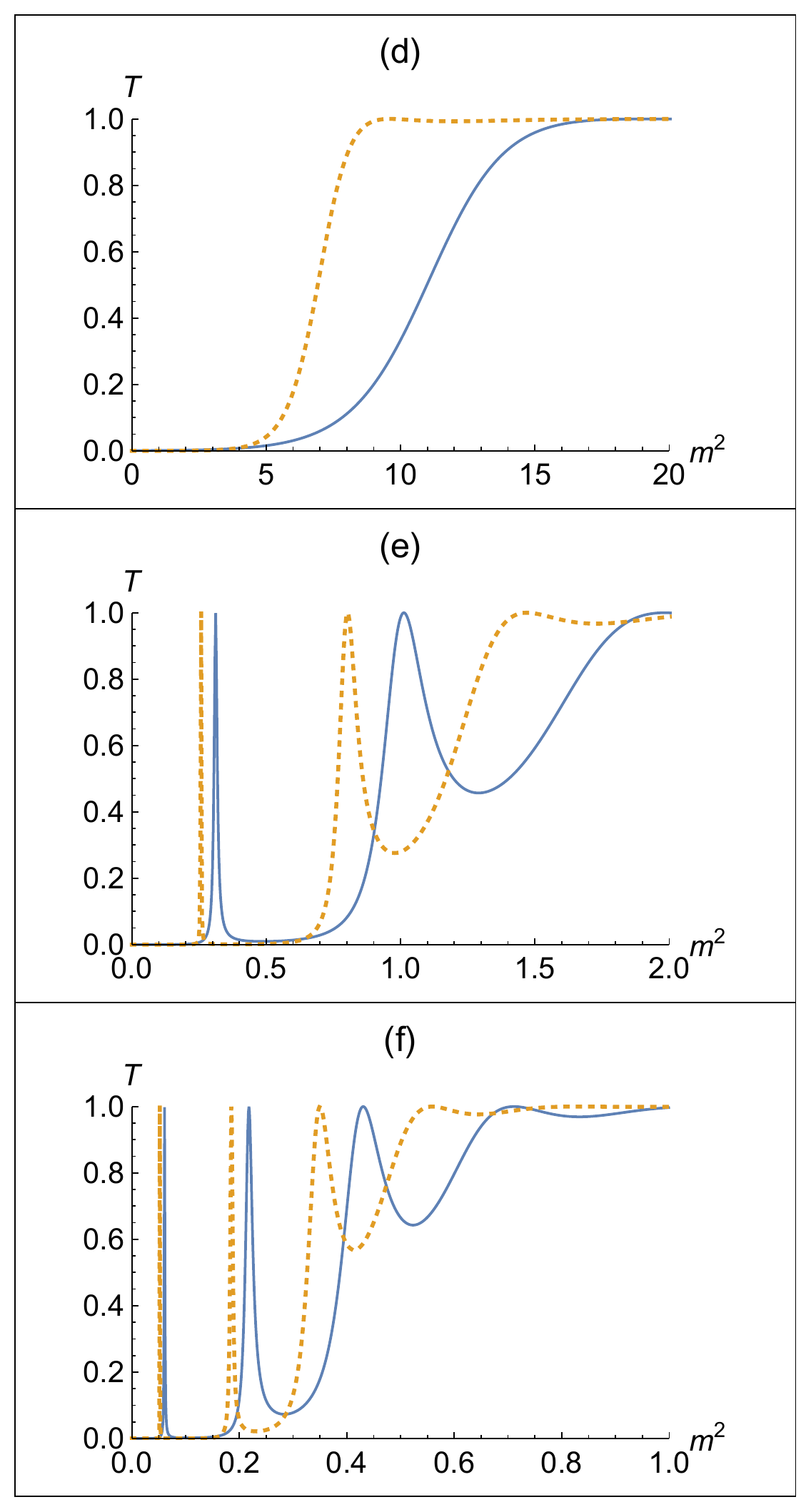}}
\caption{Potentials $U(z)$ and Transmission coefficient $T(m^2)$ for massive left fermions for $D=5$ (line) and $D=10$ (dotted), with $\eta=1$ and 
        $s=1$ (a, d), $s=3$ (b, e) and $s=5$ (c, f) in derivate dilaton setup.}
\label{Potentials6}
\end{figure}

    For right handed fermions, the potentials and transmission coefficient are showed in Fig. \ref{Potentials6}. In Figs. \ref{Potentials6}(a) and 
\ref{Potentials5}(d) we show the potential and transmission coefficient for $s=1$, when $D=5,10$. The Figs. \ref{Potentials6}(b), (c) and 
\ref{Potentials5}(e), (f) we show the graphics of potential and transmission coefficient with $s=3,5$, respectively. Here we note that the 
increase of bulk dimensionality does not contribute to the appearance of resonance peaks. Thus we conclude that only $s=3,5$ topological defects induces the appearance 
of resonance peaks for the case with derivative coupling.

\begin{table}[h]
\begin{center}
 \begin{tabular}{|c|c|c|c|c|}
  \hline
   Topological Defects    & \multicolumn{4}{|c|}{Resonances Peaks ($m^2$)} \\
  \hline
                           $s=1$ &  -    &  -  &    - &     -     \\
  \hline
                           $s=3$ &  1.6  &  -  & 0,6  &     -        \\
\hline
                           $s=5$ & 0,02 & 0,17 & 0,15 &     -        \\
 \hline
 Dimensionality         & \multicolumn{2}{|c|}{$D=5$}  & \multicolumn{2}{|c|}{$D=10$}\\
 \hline
 \end{tabular}
 \end{center}
\caption{Peaks of resonances for massive right handed fermions.}
\label{TableDilaton-2}
\end{table}

\section{Conclusion}

    In this work we have analyzed the localization of zero mode and resonances of 1/2-spin fermions in co-dimension one Randall-Sundrum-like models. This matter has 
already been studied before for many authors, but adopting the bulk with five or six dimensions. Therefore the authors generalize the study of fermions localization to consider 
the influence of space-time dimension. We discovered that the zero mode localization as well as the resonances depends strongly on this. The results of the localization of zero 
modes are given in the tables (\ref{Table odd}) and (\ref{Table even}).

\begin{table}[h]
\caption{Summary of results of zero mode fermions localization in odd dimensions. The $\psi_0(x)$ and $\psi_{0\pm}(x)$ are Dirac and Weyl spinors on the brane, respectively.}
\begin{center}
 \begin{tabular}{|c|c|c|c|}
  \hline
   Setup                                        & \multicolumn{2}{|c|}{Parameters}&  Result            \\ \hline
   \multirow{2}{*}{Thin brane}                  &  $k_p>0$                       &     -         &  $\psi_{0\pm}(x)$ is not localized \\    \cline{2-4}
                                                &  $k_p<0$                       &     -         &  $\psi_{0\pm}(x)$ localized     \\ \cline{2-4}
  \hline
   \multirow{2}{*}{Thick brane without dilaton} &    $\eta_1>0$                  &     -         &  $\psi_{0-}(x)$ is localized    \\ \cline{2-4}
                                                &    $\eta_1<0$                  &     -         &  $\psi_{0+}(x)$ is localized    \\ \cline{2-4}
  \hline
   \multirow{6}{*}{Thick brane with dilaton ($\alpha_p=0$)}   & \multirow{3}{*}{$\eta_2>0$}    & $0<r<1$     & \multirow{2}{*}{$\psi_{0-}(x)$ is localized}   \\  \cline{3-3}
                                                              &                                & $r=1$       &     \\ \cline{3-4}
                                                              &                                & $r=(p+1)/2$ & $\psi_{0\pm}(x)$ are localized     \\ \cline{2-4}
                                                              & \multirow{3}{*}{$\eta_2<0$}    & $0<r<1$     & \multirow{2}{*}{$\psi_{0+}(x)$ is localized}   \\ \cline{3-3}
                                                              &                                & $r=1$       &   \\ \cline{3-4}
                                                              &                                & $r=(p+1)/2$ & $\psi_0(x)$ is localized \\  \hline
  \multirow{7}{*}{Thick brane with dilaton ($\alpha_p\neq0$)} & \multirow{3}{*}{$\eta_2>0$}    & $0<r<1$     & \multirow{2}{*}{$\psi_{0+}(x)$ is localized if $\alpha_p<0$}\\ \cline{3-3}
                                                              &                                & $r=1$       &    \\ \cline{3-4}
                                                              &                                & $r=(p+1)/2$ & $\psi_{0\pm}$ are localized     \\ \cline{2-4}
                                                              & \multirow{4}{*}{$\eta_2<0$}    & $0<r<1$     & \multirow{2}{*}{$\psi_{0-}(x)$ is localized if $\alpha_p<0$} \\ \cline{3-3}
                                                              &                                & $r=1$       &  \\ \cline{3-4}
                                             &                                & \multirow{2}{*}{$r=(p+1)/2$} & $\psi_{0-}(x)$  is localized if $\alpha_p<0$ \\ \cline{4-4}
                                                              &                                &             & $\psi_0(x)$ is localized if $\alpha_p>0$ \\ \hline
                                                              &  \multirow{3}{*}{$h\neq 0$}    & $0<r<1$     &  $\psi_{0\pm}(x)$ is not localized \\ \cline{3-4}
    Dilaton derivate                                          &                                & $r=1$       &  $\psi_{0\pm}(x)$ is not localized \\  \cline{3-4}
                                                              &                                & $r=(p+1)/2$ &  $\psi_{0\pm}(x)$ is localized   \\
 \hline
 \end{tabular}
 \end{center}
 \label{Table odd}
\end{table}

    We  found  that in the delta-like braneworld, both the odd and even dimensions, the zero mode of fermions are not localized, 
except for negative tension brane ($k_p<0$). For thick brane without dilaton but with Yukawa coupling  $F(\phi)=\eta_1\phi$, the zero mode is localized for one chirality for odd dimensions, if the 
condition $|\eta_1|>\beta_pb/a$ is satisfied. In this situation, for odd dimensions left chirality is localized if $\eta_1>0$ or right chirality if $\eta_1<0$. However, for even 
dimensions, we discovered that Dirac spinor is localizable if $\xi_{0+}(y)=0$ and $\xi_{0-}(y)\neq0$ for $\eta_1>0$ or $\xi_{0+}(y)\neq0$ and $\xi_{0-}(y)=0$ for $\eta_1<0$. In the 
dilaton scenario, $F(\phi,\pi)=\eta_2\phi e^{-\lambda \pi}$, the zero mode localization in co-dimension one for Dirac fermions is divided into two cases: $\alpha_p=0$ and 
$\alpha_p\neq0$, whose localization depends on the parameters $\eta_2$ and $r$ to be successful (tables \ref{Table odd} and \ref{Table even}). In the case $\alpha_p=0$, for 
$\eta_2>0$ ($\eta_2<0$) and $0<r<1$ or $\eta_2>0$ ($\eta_2<0$) and $r=1$ we have that left (right) handed fermions $\psi_{0-}$ ($\psi_{0+}$) are localized 
for odd dimensions and Dirac spinor $\psi_0(x)$ is localized if $\xi_{0-}(y)\neq0$ ($\xi_{0+}(y)\neq0$) and $\xi_{0+}(y)=0$ ($\xi_{0-}(y)=0$).  In the case $\alpha_p\neq0$, right 
(left) handed fermions $\psi_{0+}(x)$ ($\psi_{0-}(x)$) are localized in the cases $\eta_2>0$ ($\eta_2<0$) $0<r<1$ and $r=1$ if $\alpha_p<0$. For $r=(p+1)/2$ Dirac spinor is 
localized independently of $\alpha_p$, both odd and even dimensions and also independent of the signal of $\eta_2$. Lastly, in the derivate dilaton coupling case, 
$H(\phi,\pi)=h\pi$, the parameter $h$ does not influence in the localization mechanism and Dirac's fermions are localizable only if $r=(p+1)/2>1$. 

\begin{table}[h]
\caption{Summary of results of zero mode fermions localization in even dimensions. The $\psi_0(x)$ is the Dirac spinor on the brane.}
\begin{center}
 \begin{tabular}{|c|c|c|c|}
  \hline
   Setup                                        & \multicolumn{2}{|c|}{Parameters}&  Result            \\ \hline
   \multirow{2}{*}{Thin brane}                  &  $k_p>0$                       &     -         &  $\psi_0(x)$ is not localized \\    \cline{2-4}
                                                &  $k_p<0$                       &     -         &  $\psi_0(x)$ localized     \\ \cline{2-4}
  \hline
   \multirow{2}{*}{Thick brane without dilaton} &    $\eta_1>0$                  &     -         &  $\psi_0(x)$ is localized if $\xi_{0+}(y)=0$   \\ \cline{2-4}
                                                &    $\eta_1<0$                  &     -         &  $\psi_0(x)$ is localized if $\xi_{0-}(y)=0$  \\ \cline{2-4}
  \hline
   \multirow{6}{*}{Thick brane with dilaton ($\alpha_p=0$)}   & \multirow{3}{*}{$\eta_2>0$}    & $0<r<1$     & \multirow{2}{*}{$\psi_0(x)$ is localized if $\xi_{0+}(y)=0$}\\ \cline{3-3}
                                                              &                                & $r=1$       &   \\ \cline{3-4}
                                                              &                                & $r=(p+1)/2$ & $\psi_0(x)$ are localized     \\ \cline{2-4}
                                                              & \multirow{3}{*}{$\eta_2<0$}    & $0<r<1$     & \multirow{2}{*}{$\psi_0(x)$ is localized if $\xi_{0-}(y)=0$} \\ \cline{3-3}
                                                              &                                & $r=1$       &  \\ \cline{3-4}
                                                              &                                & $r=(p+1)/2$ & $\psi_0(x)$ is localized \\  \hline
  \multirow{7}{*}{Thick brane with dilaton ($\alpha_p\neq0$)} & \multirow{3}{*}{$\eta_2>0$}    & $0<r<1$     & $\psi_0(x)$ is localized if $\alpha_p<0$ \\  \cline{3-3}               
                                                              &                                & $r=1$       & and $\xi_{0-}(y)=0$   \\ \cline{3-4}
                                                              &                                & $r=(p+1)/2$ & $\psi_0(x)$ are localized     \\ \cline{2-4}
                                                              & \multirow{4}{*}{$\eta_2<0$}    & $0<r<1$     & $\psi_0(x)$ is localized if $\alpha_p<0$  \\ \cline{3-3}
                                                              &                                & $r=1$       & and $\xi_{0+}(y)=0$ \\ \cline{3-4}
                                             &                                & \multirow{2}{*}{$r=(p+1)/2$} & $\psi_{0}(x)$  is localized if $\alpha_p<0$ \\ 
                                             &                                &                              & and $\xi_{0+}(y)=0$ \\ \cline{4-4}
                                                              &                                &             & $\psi_0(x)$ is localized if $\alpha_p>0$                                                     \\ \hline
                                                              &  \multirow{3}{*}{$h\neq 0$}    & $0<r<1$     &  $\psi_0(x)$ is not localized \\ \cline{3-4}
    Dilaton derivate                                          &                                & $r=1$       &  $\psi_0(x)$ is not localized \\  \cline{3-4}
                                                              &                                & $r=(p+1)/2$ &  $\psi_0(x)$ is localized   \\
 \hline
 \end{tabular}
 \end{center}
 \label{Table even}
\end{table}

    For the study of resonances in undeformed and deformed Randall-Sundrum Models, we used the transfer matrix method. In all Randall-Sundrum-like models the issue 
of localization is studied through an associated one dimensional Schr\"odinger equation with a respective potential. In general, the results obtained enforces the 
previous affirmation that the resonances are strongly dependent on value of parameter $\eta_2$ and $h$. Beyond this, both cases are strongly dependent on $s$. 
The authors of this paper also observed that, when $\lambda=1/2\sqrt{pM^p}$, specifically for $D=10$ ($p=8$), the resonance peaks are dependent of the bulk 
dimensionality, as shown in Fig. \ref{Potentials1}(d). However, the same does not occur for the cases with $s=3,5$, where these topological defects are responsible
for the appearance of resonance peaks. Another interesting aspect is the displacement of the resonance peaks with the increase of the bulk dimensionality. This 
happens in the Figs. ($s=5$, Figs. \ref{Potentials2}(f) and \ref{Potentials3}(f)) and Figs. ($s=3,5$, Figs. \ref{Potentials1}(e) and (f)).
 Phenomenologically, this fact can be interpreted as an increase of the energy scale. In the derivate coupling case, for left handed fermions the inverse occurs: 
in the cases $s=3,5$, the numbers of resonance peaks diminish with increasing bulk dimensionality for left handed fermions (odd dimensions) and for Dirac fermions 
(even dimensions, choosing the function $\xi_{-n}(y)$) (see Figs. \ref{Potentials5}(e) and (f)). The same is not true for right handed fermions (odd dimensions) or
for Dirac Fermions (even dimensions choosing $\xi_{+n}(y)$). 

    Finally, we can conclude that both the localization of the zero mode and the appearance of unstable resonant modes are affected by space-time dimensionality. For
example, we note the possibility of localizing both chiralities and the appearance of new peaks and shifts of these peaks in domain walls when the dimensionality 
increase. We can summarize the main results of this work as follows:
\begin{enumerate}
 \item For even dimensions the zero mode of Dirac spinor can be localized, prohibiting massive modes.
 \item For odd dimensions only one zero mode chiralities can be localized.
 \item The localization of zero mode depends strongly on the coupling constants and the parameter $r$.
 \item The resonance peaks for massive modes are displaced with increasing dimensionality.
\end{enumerate}

\subsection*{Acknowledgments}

    We acknowledge the physics departments of the Universidade Federal do Cear\'a (UFC), Universidade Estadual do Cear\'a (UECE), the financial support provided by 
Conselho Nacional de Desenvolvimento Cient\'{\i}fico e Tecnol\'ogico (CNPq) and Funda\c c\~ao Cearense de 
Apoio ao Desenvolvimento Cient\'\i fico e Tecnol\'ogico (FUNCAP)  through PRONEM PNE-0112-00085.01.00/16.
\label{End}

\providecommand{\href}[2]{#2}\begingroup\raggedright\endgroup

%
\end{document}